\documentclass[journal,12pt,onecolumn,draftclsnofoot]{IEEEtran}

\usepackage{xcolor}
\usepackage{graphicx}
\usepackage{epstopdf}
\usepackage{subfigure}
\usepackage{amsmath}
\usepackage{amsthm}
\usepackage{amssymb}
\newtheorem{theorem}{Theorem}
\newtheorem{lemma}[theorem]{Lemma}
\newtheorem{example}{Example}
\newtheorem{definition}{Definition}
\usepackage{amsfonts}
\usepackage{mathdots}
\usepackage{algorithm}
\usepackage[noend]{algorithmic}
\usepackage[noadjust]{cite}
\usepackage[acronym,nogroupskip,nonumberlist,nopostdot]{glossaries}
\usepackage{xspace}
\usepackage{mfirstuc}

\newacronym{ser}{SER}{symbol error rate}
\newcommand{\ser}{\gls{ser}\xspace}

\newacronym{snr}{SNR}{signal-to-noise ratio}
\newcommand{\snr}{\gls{snr}\xspace}

\newacronym{qpsk}{QPSK}{quadrature phase shift keying}
\newacronym{mpsk}{M-PSK}{M-phase shift keying}

\newacronym{smi}{SMI}{secrecy mutual information}
\newcommand{\smi}{\gls{smi}\xspace}

\newacronym{mimo}{MIMO}{multiple-input multiple-output}
\newcommand{\mimo}{\gls{mimo}\xspace}

\newacronym{dm}{DM}{directional modulation}
\newcommand{\dm}{\gls{dm}\xspace}
\newcommand{\DM}{\Gls{dm}\xspace}

\newacronym{AN}{AN}{artificial noise}
\newcommand{\AN}{\gls{AN}\xspace}

\newacronym{APN}{APN}{artificial phase noise}
\newcommand{\apn}{\gls{APN}\xspace}

\newacronym{rf}{RF}{radio frequency}
\newcommand{\rf}{\gls{rf}\xspace}

\newacronym{pls}{PLS}{physical layer security}
\newcommand{\pls}{\gls{pls}\xspace}

\newcommand{\imj}{\mathsf{j}}
\newcommand{\Ntx}{N_{\mathrm{T}}}

\newcommand{\MU}{\mathrm{R}}
\newcommand{\Eve}{\mathrm{E}}

\newcommand{\rSub}[2]{\ifthenelse{\equal{#2}{}}
    {r_{\mathrm{#1}}}
    {r_{\mathrm{#1}, {#2}}}}
\newcommand{\thetaSub}[2]{\ifthenelse{\equal{#2}{}}
    {\theta_{\mathrm{#1}}}
    {\theta_{\mathrm{#1}, {#2}}}}
\newcommand{\phiSub}[2]{\ifthenelse{\equal{#2}{}}
    {\phi_{\mathrm{#1}}}
    {\phi_{\mathrm{#1}, {#2}}}}
\newcommand{\iSub}[2]{\ifthenelse{\equal{#2}{}}
    {i_{\mathrm{#1}}}
    {i_{\mathrm{#1}, {#2}}}}
\newcommand{\jSub}[2]{\ifthenelse{\equal{#2}{}}
    {j_{\mathrm{#1}}}
    {j_{\mathrm{#1}, {#2}}}}

\newcommand{\rMU}[1][]{\ifthenelse{\equal{#1}{}}
    {\rSub{\MU}{}}
    {\rSub{\MU}{#1}}}
\newcommand{\thetaMU}[1][]{\ifthenelse{\equal{#1}{}}
    {\thetaSub{\MU}{}}
    {\thetaSub{\MU}{#1}}}
\newcommand{\phiMU}[1][]{\ifthenelse{\equal{#1}{}}
    {\phiSub{\MU}{}}
    {\phiSub{\MU}{#1}}}
\newcommand{\iMU}[1][]{\ifthenelse{\equal{#1}{}}
    {\iSub{\MU}{}}
    {\iSub{\MU}{#1}}}
\newcommand{\jMU}[1][]{\ifthenelse{\equal{#1}{}}
    {\jSub{\MU}{}}
    {\jSub{\MU}{#1}}}
\newcommand{\rEve}[1][]{\ifthenelse{\equal{#1}{}}
    {\rSub{\Eve}{}}
    {\rSub{\Eve}{#1}}}
\newcommand{\thetaEve}[1][]{\ifthenelse{\equal{#1}{}}
    {\thetaSub{\Eve}{}}
    {\thetaSub{\Eve}{#1}}}
\newcommand{\phiEve}[1][]{\ifthenelse{\equal{#1}{}}
    {\phiSub{\Eve}{}}
    {\phiSub{\Eve}{#1}}}
\newcommand{\iEve}[1][]{\ifthenelse{\equal{#1}{}}
    {\iSub{\Eve}{}}
    {\iSub{\Eve}{#1}}}
\newcommand{\jEve}[1][]{\ifthenelse{\equal{#1}{}}
    {\jSub{\Eve}{}}
    {\jSub{\Eve}{#1}}}

\newcommand{\thetatilt}{\thetaSub{\mathrm{tilt}}{}}

\newcommand{\Ts}{T_\mathrm{s}}
\newcommand{\vmax}{v_\mathrm{max}}

\newcommand{\set}[1]{\left[#1\right]}
\newcommand{\modulo}[2][\Ntx]{\left(#2\right)_{\%#1}}
\newcommand{\frob}[2]{\mathchoice{\left\langle#1,#2\right\rangle}{\langle#1,#2\rangle}{\langle#1,#2\rangle}{\langle#1,#2\rangle}}
\newcommand{\Ph}{\measuredangle}
\DeclareMathOperator*{\argmin}{arg\,min}
\DeclareMathOperator*{\argmax}{arg\,max}

\newcommand{\figref}[1]{Fig. \ref{#1}}
\newcommand{\algref}[1]{Algorithm \ref{#1}}
\newcommand{\secref}[1]{Section \ref{#1}}
\newcommand{\lemmaref}[1]{Lemma \ref{#1}}
\newcommand{\defref}[1]{Definition \ref{#1}}
\newcommand{\exref}[1]{Example \ref{#1}}

\makeglossaries

\makeatletter
\if@twocolumn
\newcommand{\iftwocolumn}[1]{#1}
\else
\newcommand{\iftwocolumn}[1]{}
\fi
\makeatother

\allowdisplaybreaks
\begin{document}
	\title{Circulant Shift-based Beamforming\\ for Secure Communication with \\Low-resolution Phased Arrays}
	\author{{\IEEEauthorblockN{Kartik Patel, Nitin Jonathan Myers, Robert W. Heath Jr.}}
		\thanks{K. Patel is with the Wireless Networking and Communications Group, The University of Texas at Austin, USA. N.J. Myers is with the Delft Center for Systems and Control, Delft University of Technology, The Netherlands. R.W. Heath Jr. is with the Department of Electrical and Computer Engineering, North Carolina State University, USA. This material is based upon work supported in part by the National Science Foundation under the grant number CNS-1731658 and the Army Research Office under grant W911NF1910221.}}
	\maketitle
	\begin{abstract}
		Millimeter wave (mmWave) technology can achieve high-speed communication due to the large available spectrum. Furthermore, the use of directional beams in mmWave system provides a natural defense against physical layer security attacks. In practice, however, the beams are imperfect due to mmWave hardware limitations such as the low-resolution of the phase shifters. These imperfections in the beam pattern introduce an energy leakage that can be exploited by an eavesdropper. To defend against such eavesdropping attacks, we propose a directional modulation-based defense technique where the transmitter applies random circulant shifts of a beamformer. We show that the use of random circulant shifts together with appropriate phase adjustment induces \apn in the directions different from that of the target receiver. Our method corrupts the phase at the eavesdropper without affecting the communication link of the target receiver. We also experimentally verify the \apn induced due to circulant shifts, using channel measurements from a 2-bit mmWave phased array testbed. Using simulations, we study the performance of the proposed defense technique against a greedy eavesdropping strategy in a vehicle-to-infrastructure scenario. The proposed technique achieves better defense than the antenna subset modulation, without compromising on the communication link with the target receiver.
		
	\end{abstract}
	\IEEEpeerreviewmaketitle
	
	\section{Introduction}
	Millimeter wave (mmWave) communication uses directional beamforming where signals are transmitted or received along selected directions~\cite{HeaGonRan:An-Overview-of-Signal-Processing:16}. 
	Directional beamforming also provides resilience against eavesdropping attacks as it concentrates the transmitted \rf signals along the direction of the intended user and reduces the signal transmitted along \textit{unintended} directions, i.e. directions other than the direction of the intended user~\cite{WuKhiXia:A-Survey-of-Physical-Layer:18}. 
	
	The directional beam patterns, in practice, are not perfect due to the design constraints in mmWave radios. Due to the high power consumption with fully digital arrays in a wideband setting, commodity mmWave radios are usually based on hybrid or analog antenna arrays that use RF phase shifters~\cite{HeaGonRan:An-Overview-of-Signal-Processing:16}. Moreover, the resolution of the RF phase shifters in these arrays is limited to few bits to reduce the hardware complexity~\cite{PooTag:Supporting-and-Enabling-Circuits:12}. The low resolution of phase shifters results in imperfections in the directed beam patterns which \textit{leak} the RF signal along the unintended directions. In this paper, we study the RF signals leaked with such low resolution phased arrays and show that this leakage can be exploited by a mobile eavesdropper, such as an unmanned aerial vehicle (UAV) in a vehicle-to-infrastructure (V2I) scenario.
	
	\par A standard approach to improve \pls in an mmWave system is to reduce the energy leakage by appropriately designing a beamformer using channel state information (CSI) or the position of the eavesdropper~\cite{LeeChoLee:Exploiting-Array-Pattern:19, JuWanZhe:Secure-Transmissions-in-Millimeter:17}. In \cite{LeeChoLee:Exploiting-Array-Pattern:19}, a precoding technique was proposed to reduce the energy leaked along the direction of the eavesdropper. In \cite{JuWanZhe:Secure-Transmissions-in-Millimeter:17}, defense mechanisms that exploit partial CSI to design precoders were developed to minimize the energy leakage. 
	In this work, we claim that an eavesdropper can still breach such defenses that only focus on minimizing the energy leakage along potential eavesdropping directions. This is because a mobile eavesdropper can still achieve good received power by moving to a different direction, or by shifting closer to the transmitter (TX). The defense techniques in~\cite{LeeChoLee:Exploiting-Array-Pattern:19} and~\cite{JuWanZhe:Secure-Transmissions-in-Millimeter:17} also require fully digital antenna arrays and partial information about the eavesdropper, neither of which may be available in a practical system with analog or hybrid phased arrays. 
	
	\par Defense mechanisms that do not require fully digital arrays and are unaware of the eavesdropping location were proposed in~\cite{TiaLiWan:Hybrid-Precoder-and-Combiner:17,ZhuWanWon:Secure-Communications-in-Millimeter:17}. 
	In \cite{TiaLiWan:Hybrid-Precoder-and-Combiner:17, ZhuWanWon:Secure-Communications-in-Millimeter:17}, hybrid beamformers were designed to transmit \AN along the unintended directions. 
	Such \AN-based defense techniques, however, degrade the performance at the intended receiver (RX). This is because either \AN is induced at the RX or the power allocated for data transmission is reduced. An alternative approach that induces spatially selective \AN requires partial CSI or position information of the eavesdropper which may not be available at the TX~\cite{LiMa:Spatially-Selective-Artificial-Noise:13,LiaChaMa:QoS-Based-Transmit-Beamforming:11,GerSchJor:Secrecy-Outage-in-MISO:12}. 
	
	\par \DM-based physical layer defense techniques are also promising for secure mmWave communication. These methods modify the beamformer at every symbol such that the constellation is maintained along the intended direction and distorted along other directions~\cite{DinFus:A-Vector-Approach-for-the-Analysis:14,HafYusKha:Secure-Spatial-Multiple:18,KalSolMal:Directional-Modulation-Via-Symbol-Level:16,KalSolMal:Secure-M-PSK-communication:16,ShuXuWan:Artificial-Noise-Aided-Secure-Multicast:18,ChrBor:Iterative-Convex-Optimization:18,ValLozHea:Antenna-Subset-Modulation:13,EltChoAl-:On-the-Security-of-Millimeter-Wave:16, WanZhe:Hybrid-MIMO-and-Phased-Array:18,WeiMasLiu:Secure-directional-modulation:2021,DalBer:Directional-Modulation-Technique:09}. 
	Various algorithms to design \dm-based symbol-level precoding have been proposed for secure \mimo communication with a digital antenna array~\cite{DinFus:A-Vector-Approach-for-the-Analysis:14,HafYusKha:Secure-Spatial-Multiple:18,KalSolMal:Directional-Modulation-Via-Symbol-Level:16,KalSolMal:Secure-M-PSK-communication:16,ShuXuWan:Artificial-Noise-Aided-Secure-Multicast:18,ChrBor:Iterative-Convex-Optimization:18}. 
	In the context of mmWave systems with hybrid or analog antenna array, \dm-based methods have been proposed in \cite{DalBer:Directional-Modulation-Technique:09,ValLozHea:Antenna-Subset-Modulation:13, EltChoAl-:On-the-Security-of-Millimeter-Wave:16,WanZhe:Hybrid-MIMO-and-Phased-Array:18,WeiMasLiu:Secure-directional-modulation:2021}. 
	For instance, the Antenna Subset Modulation (ASM) technique proposed in \cite{ValLozHea:Antenna-Subset-Modulation:13} switches off a subset of antennas at every symbol. Switching at random changes the beamformer which affects the amplitude and phase of the transmitted symbol in all directions. By adjusting the phase of the transmitted symbol, the intended symbol is received at the RX while the symbol at the eavesdropper is distorted. 
	A similar technique in \cite{EltChoAl-:On-the-Security-of-Millimeter-Wave:16} selects a random subset of antennas to destructively combine the RF signals at the unintended directions. Unfortunately, the methods in \cite{ValLozHea:Antenna-Subset-Modulation:13, EltChoAl-:On-the-Security-of-Millimeter-Wave:16} reduce the mainlobe gain under the per-antenna power constraint. As a result, the RX observes a lower power when compared to the use of an ideal directional beam. 
	In \cite{WanZhe:Hybrid-MIMO-and-Phased-Array:18}, a time-modulated \dm-based technique was proposed for secure mmWave communication. Another \dm-based technique for \textit{actively driven} phased arrays, where an amplifier is cascaded after each low-resolution phase shifter, was developed in \cite{WeiMasLiu:Secure-directional-modulation:2021}. Our defense technique, in contrast, is designed for low-resolution phased arrays with passive phase shifters under the per-antenna power constraint. Our method also does not require CSI of the eavesdropper.
	
	\par In this paper, we propose a novel \dm-based approach to defend against an eavesdropper without impacting the communication performance at the RX. Our method called \textit{Circulant Shift-based Beamforming} (CSB) applies a random circulant shift of the standard beamformer in every symbol duration. These random circulant shifts induce random phase changes in the symbols received along different directions. As the TX knows the phase change induced along the intended direction, it adjusts the transmitted symbol such that the RX receives the symbol without any phase distortion. The symbol observed along any other direction, however, is corrupted by \apn. We characterize the statistical properties of the \apn induced by CSB along the on-grid directions and show that the equivalent channel between the TX and the eavesdropper suffers from an ambiguity in the phase of the received symbol. As a result, coherent modulation techniques such as $M$-PSK cannot be decoded by an eavesdropper located along the on-grid directions even if the eavesdropper observes a high received power.
	
	\par The proposed CSB has three key advantages over the techniques designed for mmWave systems. 
	First, there is a smaller power loss at the RX compared to the ASM-based approach, as CSB activates all the antennas. Furthermore, circulantly shifting a beamformer does not change the beamforming gain at the discrete angles defined by the common DFT codebook. 
	Second, our method is designed for low-resolution phased arrays without the assumption of active antenna elements as opposed to the prior work in~\cite{WeiMasLiu:Secure-directional-modulation:2021}.  
	Third, CSB has a low complexity than other \dm-based beamforming methods as CSB does not require any real-time optimization to compute the beamformer to achieve secure communication. 
	
	\par We would like to mention that our technique is different from recent \pls methods based on spatial modulation (SM) ~\cite{WanWanSun:Spatial-modulation-aided:16} and index modulation (IM)~\cite{LeeJoKo:Secure-Index-and-Data:17}. 
	In the SM-based defense techniques~\cite{WanWanSun:Spatial-modulation-aided:16}, the TX selects a subset of antennas based on the CSI of the channel between itself and the RX. Then, the RX uses the CSI to decode the data symbols. 
	An IM-based defense technique such as the one discussed in \cite{LeeJoKo:Secure-Index-and-Data:17} uses rule-based mapping for index modulation in OFDM-IM. 
	In contrast, our proposed CSB defense does not focus on antenna selection or IM. Our method only applies circulant shifts of the beamformer to corrupt the phase of the received symbols at the eavesdropper. The contributions of this paper can be summarized as follows:
	\begin{itemize}
		\item We propose CSB for secure communication under RF energy leakage due to low resolution phase shifters. 
		Our technique applies random circulant shifts of the beamformer together with appropriate phase correction in the transmitted symbol, to introduce \apn in the unintended directions. The phase correction ensures that the RX obtains the correct transmitted symbol. We also theoretically analyze the \smi of the proposed defense technique.
		\item We validate the key idea underlying the proposed defense mechanism using an mmWave phased array testbed. Considering the phase noise limitation of our phased arrays, we design an experiment suitable to measure the phase change induced due to circulant shifts. We, then, experimentally show that circulant shifts induce different phase shifts along different directions.
		\item We design a first of its kind mobile eavesdropping attack in a V2I mmWave system with low-resolution phased arrays. For this attack, we formulate a 2D trajectory optimization problem to track the directions of the RF energy leakage over time. We numerically show how standard beamforming is vulnerable to such an attack, and discuss the use of CSB technique to defend this attack. 
	\end{itemize}
	
	\textbf{Organization:} \secref{sec:systemModel} contains the geometrical channel model and the definitions used in the paper.
	In \secref{sec:defense}, we describe the proposed CSB for secure communication. Our experiment design to validate the proposed CSB is explained in \secref{sec:experiment}. In \secref{sec:attack}, we discuss our trajectory optimization-based mobile eavesdropping attack on the low-resolution phased array. Finally, we give simulation results in \secref{sec:numerical}.
	
	\textbf{Notations:} We denote the unit imaginary number by $\imj = \sqrt{-1}$. 
	We use boldface capital letter $\mathbf{A}$ to denote a matrix, boldface small letter $\mathbf{a}$ to denote a vector, and $a,A$ to denote scalars. 
	$\mathbf{A}^\text{T}, \bar{\mathbf{A}}$, and $\mathbf{A}^*$ represent the transpose, conjugate  and conjugate transpose of $\mathbf{A}$.
	We denote $(i,j)-$th element of the matrix $\mathbf{A}$ by $\left[\mathbf{A}\right]_{i,j}$. 
	The inner product of matrices $\mathbf{A} $ and $\mathbf{B}$ is defined as $\left\langle\mathbf{A}, \mathbf{B}\right\rangle=\sum_{i,j}\left[\mathbf{A}\right]_{i,j}\left[\bar{\mathbf{B}}\right]_{i,j}$. 
	We use $\set{N}$ to denote the set $\{0,1,...,N-1\}$.
	
	\section{System Model}\label{sec:systemModel}
	In this section, we describe the channel and the system model used in this paper. We also discuss the imperfections in the beams generated with low-resolution phased arrays.
	
	\subsection{Coordinate system}\label{sec:geometricalSetup}
	
	\begin{figure}[tb]
		\centering
		\subfigure[3D view \label{fig:SphericalConversion_a}]{
			\includegraphics[height=2in]{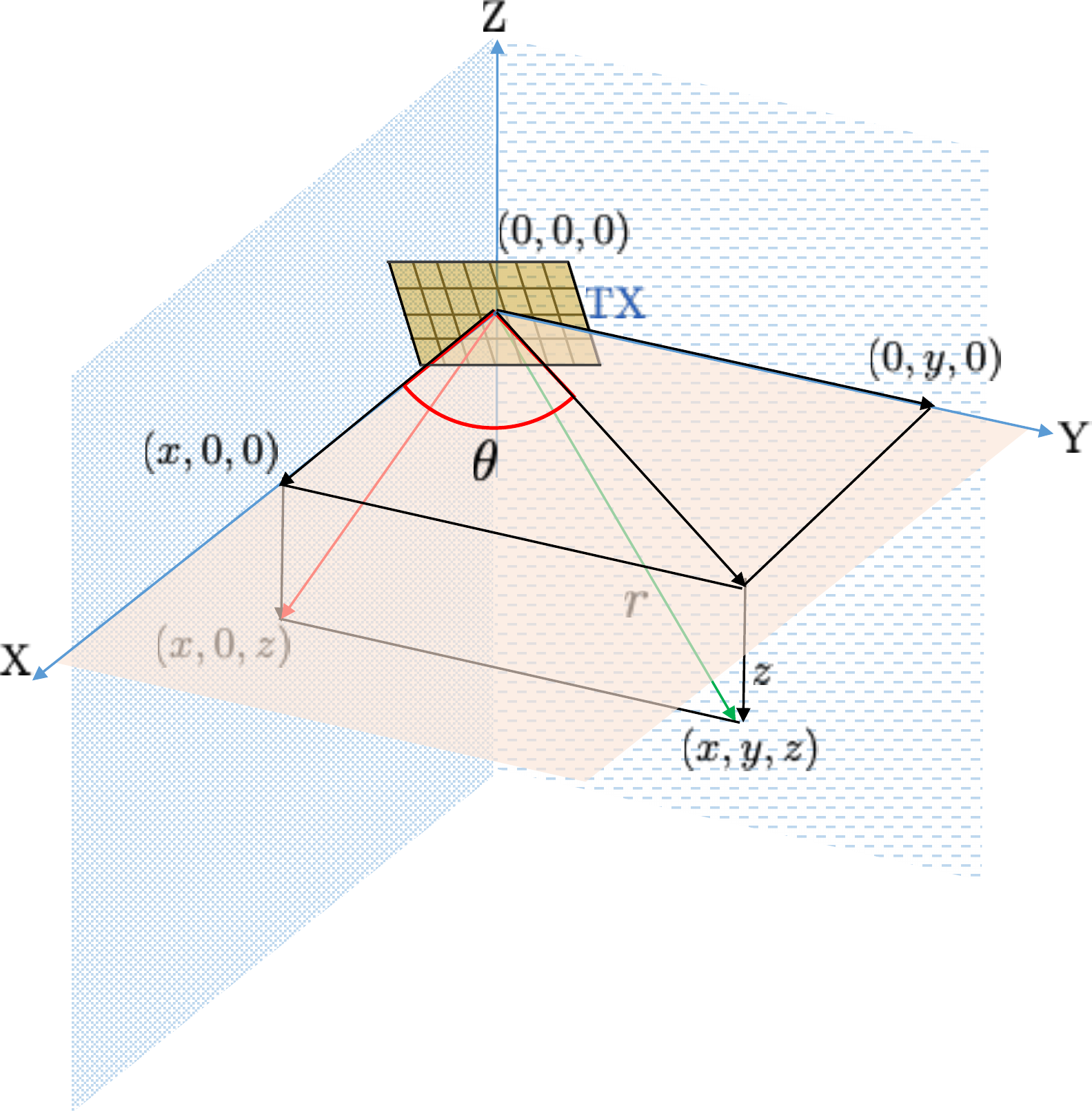}
		}\quad\quad\quad\quad
		\subfigure[\centering Side view from positive Y-axis \label{fig:SphericalConversion_b}]{
			\includegraphics[height=2in]{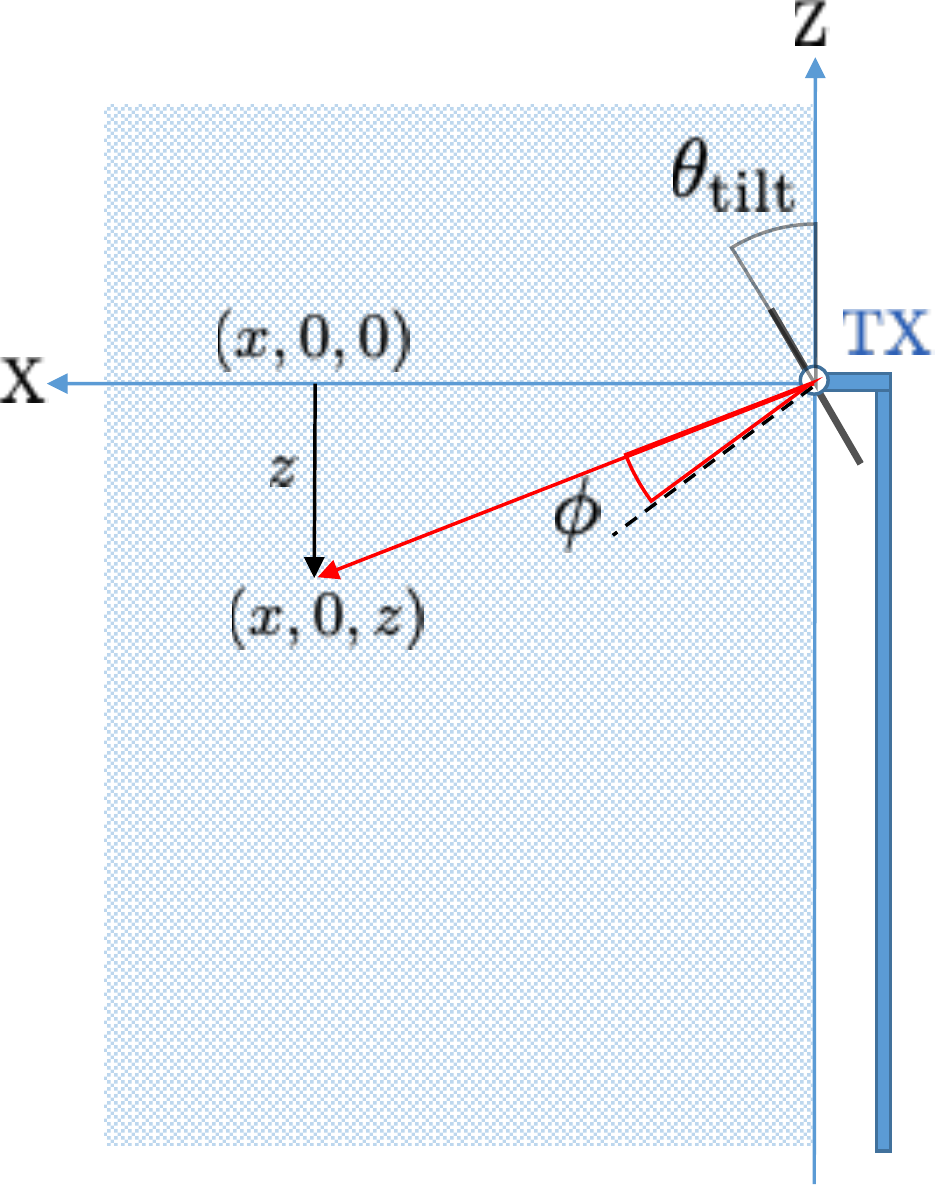}
		}
		\caption{Conversion from rectangular coordinate $(x,y,d)$ to modified spherical coordinate $(r,\theta,\phi)$. The origin of both the coordinate systems is defined as the center of the TX antenna array.}
		\label{fig:SphericalConversion}
	\end{figure}
	
	We consider the geometrical setup depicted in \figref{fig:SphericalConversion} where the TX is equipped with a planar antenna array centered at $(0,0,0)$. The plane of the TX array is perpendicular to the XZ-plane, and the array is tilted at an angle $\thetatilt$ towards the ground. For ease of analysis, we convert the rectangular coordinate system into a modified spherical coordinate system shown in \figref{fig:SphericalConversion}. The origin of the modified spherical coordinate system is defined as the center of the TX array. Consider a point $(x,y,z)$ in the rectangular coordinate system, such that, $x\geq 0$ and $y,z \in \mathbb{R}$. 
	The corresponding transformed coordinate $(r,\theta,\phi)$, where $r$ is the distance of the point from the origin, $\theta$ and $\phi$ are the azimuth and elevation angles, can be calculated as
	\begin{equation}\label{eq:modifiedSphericalSystem}
		r = \sqrt{x^2 + y^2 + z^2}, \quad \theta = \arctan \left(\frac{y}{x}\right),\quad  \phi = \arctan \left(\frac{z}{x}\right) + \thetatilt.
	\end{equation}
	We observe from the geometry that $r\geq 0$, $\theta \in [-\pi/2 ,\pi/2]$ and $\phi \in [-\pi/2+\thetatilt, \pi/2]$. For simplicity of notation, we define the mapping $S_1$ such that, $(r,\theta,\phi) = S_1((x,y,z))$.
	
	The modified spherical coordinate system defines the elevation angle as the angle between the projections of $(x,y,z)$ and the perpendicular to the TX array on XZ-plane. In contrast, the conventional spherical coordinate system defines the elevation angle as the angle between the $(x,y,z)$ and its projection on XY-plane. This modification in coordinate system simplifies the definition of the array response matrix by decoupling the phase variations across two dimensions of the TX array.
	
	\par We denote the RX coordinate in the rectangular and modified spherical systems by $(x_\MU, y_\MU, z_\MU)$ and $(r_\MU, \thetaMU, \phiMU)$. These coordinates are defined under the assumption that the center of the TX is $(0,0,0)$. Similarly, we use $(x_\Eve, y_\Eve, z_\Eve)$ and $(r_\Eve, \thetaEve, \phiEve)$ to represent the coordinates of the eavesdropper in the rectangular and the modified spherical systems. We also define the \textit{angular coordinates} of the RX and the eavesdropper, relative to the TX, as $(\thetaMU, \phiMU)$ and $(\thetaEve,\phiEve)$. 
	
	
	\subsection{Channel model}
	In this paper, we model the mmWave channel between the TX and the RX as a narrowband line-of-sight (LoS) channel. The TX is equipped with a half-wavelength spaced uniform planar array (UPA) with $\Ntx \times \Ntx$ antenna elements. Although we assume an equal number of antennas along the azimuth and the elevation dimension for notational convenience, our design can also be generalized to other rectangular array geometries. The RX and the eavesdropper are assumed to be in the far field of the TX. For simplicity, we assume that the RX and the eavesdropper are equipped with a single mmWave antenna. The techniques discussed in this paper also apply to a multi-antenna RX and a multi-antenna eavesdropper under the far field assumption. 
	
	We now describe the array response matrices at the TX for the links associated with the RX and the eavesdropper. We define the Vandermonde vector 
	\begin{align}
		\mathbf{a}(\theta) & = \left[1, e^{-\imj \pi \sin\theta}, \ldots,e^{-\imj (\Ntx-1) \pi \sin \theta}\right]^\mathrm{T}. 
	\end{align}
	As the angular coordinate of the RX relative to the TX is $(\thetaMU,\phiMU)$, the array response matrix between the TX and the RX can be expressed as 
	\begin{equation}\label{eq:channelDefMU}
		\mathbf{V}_\MU = \mathbf{V}(\thetaMU,\phiMU) =\mathbf{a}(\phiMU) \mathbf{a}^\mathrm{T}(\thetaMU).
	\end{equation}
	The definition of the elevation angle $\phiMU$ in the modified spherical system allows the use of same array response function $\mathbf{a}(\cdot)$ along both dimensions of the antenna arrays. Similar to the RX, we define the array response matrix associated with the eavesdropper as
	\begin{equation}\label{eq:channelDefEve}
		\mathbf{V}_\Eve = \mathbf{V} (\thetaEve,\phiEve)= \mathbf{a}(\phiEve) \mathbf{a}^\mathrm{T}(\thetaEve).
	\end{equation}
	Under the LoS assumption, the TX-RX and the TX-eavesdropper channels are just a scaled versions of the corresponding array response matrices. 
	
	\subsection{Signal model}
	We derive the signal model at a time instant $t$ when the RX and the eavesdropper are located at $(\rMU[t], \thetaMU[t],\phiMU[t])$ and $(\rEve[t], \thetaEve[t], \phiEve[t])$. The TX array response matrices associated with the RX and the eavesdropper are denoted by $\mathbf{V}(\thetaMU[t],\phiMU[t])$ and $\mathbf{V}(\thetaEve[t],\phiEve[t])$. The TX applies a beamformer $\mathbf{F}_{t}$ to direct its signals towards the RX. We use $x_t$ to denote the symbol transmitted by the TX. We assume that both the beamformer and the transmitted symbols are normalized, i.e. $||\mathbf{F}_{t}||_F^2 = 1$ and $\mathbb{E}[|x_t|^2] = 1$. 
	We denote the phase offset due to the propagation delay between the TX and the RX by $\nu_\MU$, the power received at the RX by $P_{\rMU[t]}$, and the independent and identically distributed (IID) complex Gaussian noise by $n_{\MU,t}\sim \mathcal{CN}(0,\sigma^2)$.
	Then, the signal received by the RX at time $t$ is 
	\begin{equation}\label{eq:signalDefMU}
		y_{\MU,t} = \sqrt{P_{\rMU[t]}}e^{\imj \nu_\MU} \frob{\mathbf{V}(\thetaMU[t],\phiMU[t]) }{ \mathbf{F}_{t}} x_t + n_{\MU,t}. 
	\end{equation}
	Similarly, let $\nu_\Eve$ be the phase offset due to the propagation delay between the TX and the eavesdropper, $P_{\rEve[t]}$ be the power received by the eavesdropper, and $n_{\Eve,t}\sim \mathcal{CN}(0,\sigma^2)$ be the IID complex Gaussian noise of the channel between the TX and the eavesdropper. Then, the signal received by an eavesdropper at $(\rEve[t], \thetaEve[t], \phiEve[t])$ is 
	\begin{equation}\label{eq:signalDefEve}
		y_{\Eve,t} = \sqrt{P_{\rEve[t]}} e^{\imj \nu_\Eve}\frob{\mathbf{V}(\thetaEve[t],\phiEve[t])}{ \mathbf{F}_{t}} x_t + n_{\Eve,t}.
	\end{equation}
	Conventional beamforming methods that are agnostic to the eavesdropper maximize the signal power at the RX. For example, $\mathbf{F}_{t}=\mathbf{V}(\thetaMU[t],\phiMU[t])/ \Ntx$ results in the maximum \snr of $\rho_{\MU,t} = P_{\rMU[t]} \Ntx^2 /\sigma^2$ at the RX. Such a beamformer, however, cannot be applied in low resolution phased arrays due to the limited resolution of phase shifters. This is because the phase of the entries in $\mathbf{V}(\thetaMU[t],\phiMU[t])$ do not necessarily take quantized values. 
	
	\subsection{Practical beamformer design}\label{sec:practicalBeamformer}
	We assume that the resolution of the phase shifters is $q$ bits. In practice, $q$ is a small number to limit the hardware complexity, e.g., $1\leq q \leq 3$~\cite{Che:Hybrid-Beamforming-With:17,SelSulAls:Hybrid-Precoding-Beamforming-Design:17}. In this case, the entries of the beamformer $\mathbf{F}_t$ can only take finite phase values within the set $\mathcal{B}_q = \{\frac{2\pi i}{2^q}:i = 0, 1, \ldots, 2^q-1\}$. Under this constraint, the phase of every element in the desired unquantized beamforming matrix is usually quantized to $q$ levels for hardware compatibility. In this section, we describe the phase quantization procedure and its impact on the generated beam pattern. 
	
	The $q$-bit phase quantization function rounds the phase to the nearest element in $\mathcal{B}_q$, i.e.,
	$Q_q(x) = \argmin_{\beta \in \mathcal{B}_q} |\beta-x|.$ 
	We denote the phase of a complex number $x$ as $\Ph(x)$. Thus, we can write the $q$-bit quantized beamformer corresponding to $\mathbf{F}_t$ as 
	\begin{equation}\label{eq:quantizedBeamformer}
		\left[\tilde{\mathbf{F}}_t\right]_{k,\ell} = \exp\left\{\imj Q_q\left(\Ph\left(\left[\mathbf{F}_t\right]_{k,\ell} \right)\right)\right\}/ \Ntx.
	\end{equation}
	We would like to mention that this approach of rounding off the phase to the nearest element in $\mathcal{B}_q$ is one of many ways to calculate limited-resolution beamformer. Other methods to find the feasible beamformer are presented in \cite{Che:Hybrid-Beamforming-With:17,SelSulAls:Hybrid-Precoding-Beamforming-Design:17,WanLiLiu:Hybrid-Precoder-and-Combiner:18}.
	
\begin{figure}[tb]
		\centering
		\includegraphics[height=1.5in,width=\columnwidth,keepaspectratio]{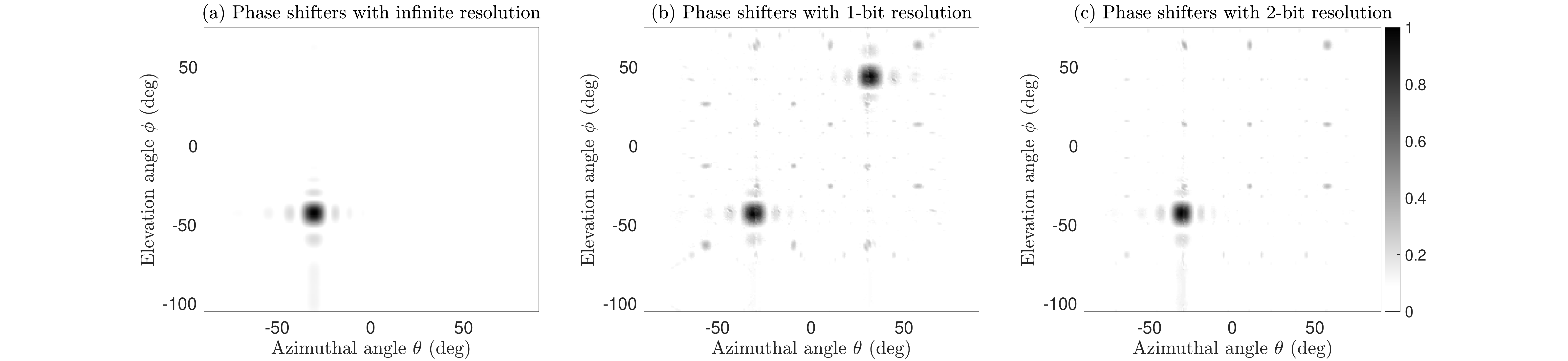}
		\caption{The normalized amplitude of the received signal in the $(\theta, \phi)$ space when beamforming is perfomed with (a) infinite-bit (b) 1-bit and (c) 2-bit resolution phased arrays. Here, the TX is equipped with a $16 \times 16$ half-wavelength spaced planar array. The array is tilted at $15^\circ$ towards ground. The TX beamforms towards an RX whose angular coordinate is $(-30^\circ, -42^\circ)$.} 
		\label{fig:energyLeakage}
	\end{figure}

	The quantization of the phase shifts introduces imperfections in the generated beam pattern.
	These imperfections cause energy leakage along the unintended directions, as shown in \figref{fig:energyLeakage}. We observe from \figref{fig:energyLeakage} that the energy leakage is significant with low-resolution phased arrays using $q=1$. Specifically, the beam patterns generated by one-bit phased arrays with a rectangular array geometry are mirror symmetric about the boresight direction (see Appendix \ref{appendix:onebitreflection} for proof). 
	
	An eavesdropper such as a mobile adversary can exploit the energy leakage by moving to the directions where the leakage is large, to eavesdrop on the TX. Furthermore, the eavesdropper can shift closer to the TX along this direction to receive a higher \snr. As a result, defense mechanisms that just minimize the energy leakage are not well suited in a mobile setting where the eavesdropper can re-position itself. Therefore, in this work, we propose a \dm-based defense mechanism that corrupts the phase of the received symbols at the eavesdropper.  Furthermore, the phase corruption due to our method is independent of the energy received by the eavesdropper. 
	
	\section{Circulant shift-based beamformer design}\label{sec:defense}
	In this section, we propose CSB as a defense against eavesdropping on a TX equipped with a low-resolution phased array. 
	\subsection{Baseline 2D-DFT codebook}
	Our CSB technique is applied on top of the standard 2D-DFT codebook used in uniform planar phased arrays. Due to the use of $q$-bit phase shifters, we define the quantized version of the 2D-DFT codebook as 
	\begin{align}
		\tilde{\mathcal{F}} & = \Bigl\{\tilde{\mathbf{F}}_{i,j}: \left[\tilde{\mathbf{F}}_{i,j}\right]_{k,\ell} = \exp\left({\imj Q_q\left(\frac{2\pi}{\Ntx}(i k+ j \ell)\right)}\right)/ \Ntx, \forall i,j,k,\ell \in \set{\Ntx}\Bigr\}.
	\end{align}
	When a beamformer is selected from the codebook $\tilde{\mathcal{F}}$, the received signal at the RX and the eavesdropper can be computed from \eqref{eq:signalDefMU} and \eqref{eq:signalDefEve}. 
	
	In the design of our defense mechanism, we assume that the RX and the eavesdropper are on-grid, i.e. $\frac{\Ntx\sin \theta}{2}$ is an integer $\forall\theta\in\{\thetaMU[t],\phiMU[t],\thetaEve[t],\phiMU[t]\}$. Although this assumption is required in the analysis of the proposed defense mechanism, we show in \secref{sec:numerical} that our method works well even when the RX is off-grid provided the angular coordinate of the RX is known. 
	
	\subsection{Circulantly shifting a beamformer}
	We define a matrix operator $\mathcal{P}_{m,n}$ that circularly shifts the input matrix by $m$ steps along each column, and by $n$ steps along every row. Specifically, for an $N\times N$ matrix $\mathbf{A}$, 
	\begin{align}
		\left[\mathcal{P}_{m,n}(\mathbf{A})\right]_{k,l} = \left[\mathbf{A}\right]_{\modulo[N]{k-m}, \modulo[N]{l-n}},
	\end{align}
	where $\modulo[N]{\cdot}$ denotes the modulo-$N$ operation. The matrix $\mathcal{P}_{m,n}(\mathbf{A})$ is interpreted as an $(m,n)$ 2D-circulant shifted version of $\mathbf{A}$.
	
	Now, we study the impact of circulantly shifting a beamformer on the received signal. We observe from \eqref{eq:signalDefMU} and \eqref{eq:signalDefEve} that the scaling introduced by the beamformer in the received symbol is $\langle \mathbf{V}(\theta,\phi), \mathbf{F} \rangle$. We define $\tilde{\mathcal{F}}$ as the set containing the $q-$bit quantized versions of the standard 2D-DFT beamformers. Our CSB technique is based on the key idea that circulantly shifting a beamformer at the TX affects the phase of the received signal differently in distinct directions. We discuss this property in Lemma \ref{lemma:lemma1}. 
	\begin{lemma}\label{lemma:lemma1}
		Let the angular coordinate of an on-grid receiver (RX or eavesdropper) be $(\theta,\phi)$ such that $\frac{\Ntx}{2}\sin \theta = i$ and $\frac{\Ntx}{2}\sin \phi = j$. If $\tilde{\mathbf{F}} \in \tilde{\mathcal{F}}$, then for any integer pair $(m,n) \in \set{\Ntx}^2$,
		\begin{equation}
			\frob{\mathbf{V}(\theta,\phi)}{ \mathcal{P}_{m,n}(\tilde{\mathbf{F}})} = \frob{\mathbf{V}(\theta,\phi)}{ \tilde{\mathbf{F}}}  e^{-\imj \frac{2\pi}{\Ntx} (m j + n i)}
		\end{equation}
		\begin{proof}
			Recall that $\mathbf{V}(\theta,\phi) = \mathbf{a}(\phi) \mathbf{a}^\mathrm{T}(\theta)$. For an on-grid receiver, the $(k,\ell)$-th element of the array response matrix $\mathbf{V}(\theta,\phi)$ is $\left[\mathbf{V}(\theta,\phi)\right]_{k,\ell} = \frac{1}{\Ntx} e^{-\imj \frac{2\pi}{\Ntx}(i k+j\ell)}$.
			In this case,
			\begin{align}
				\left\langle\mathbf{V}(\theta,\phi), \tilde{\mathbf{F}}\right\rangle &= \sum_{k,\ell} \left[\mathbf{V}(\theta,\phi)\right]_{k,\ell} \left[\tilde{\mathbf{F}}\right]_{k,\ell} =  \frac{1}{\Ntx} \sum_{k,\ell} \left[\tilde{\mathbf{F}}\right]_{k,\ell} e^{-\imj \frac{2\pi}{\Ntx} (i k + j \ell)}.
			\end{align}
			Similarly, the inner product between the circulantly shifted beamformer $\mathcal{P}_{m,n}(\tilde{\mathbf{F}})$ and $\mathbf{V}(\theta,\phi)$ is
			\begin{align}
				\frob{\mathbf{V}(\theta,\phi)}{\mathcal{P}_{m,n}(\tilde{\mathbf{F}})} &=\sum_{k,\ell} \left[\mathbf{V}(\theta,\phi)\right]_{k,\ell} \left[\mathcal{P}_{m,n}(\tilde{\mathbf{F}})\right]_{k,\ell}, \\
				&= \frac{1}{\Ntx} \sum_{k,\ell} e^{-\imj \frac{2\pi}{\Ntx} (i k + j\ell)}\left[\tilde{\mathbf{F}}\right]_{\modulo[N]{k-m}, \modulo[N]{\ell-n}}, \\
				&\stackrel{(a)}{=} \frac{1}{\Ntx} \sum_{k',\ell'} e^{-\imj \frac{2\pi}{\Ntx} (i (k'+m) + j(\ell'+n))}\left[\tilde{\mathbf{F}}\right]_{k', \ell'}= e^{-\imj \frac{2\pi}{\Ntx}(mi+nj)} \left\langle\mathbf{V}(\theta,\phi), \tilde{\mathbf{F}}\right\rangle.
			\end{align}
			where $(a)$ is based on the observation $k' = \modulo{k-m}$ and $\ell' = \modulo{\ell-n}$. 
		\end{proof}
	\end{lemma}
	
	We make three key observations from our result in \lemmaref{lemma:lemma1}. First, as $|\langle \mathbf{V}(\theta,\phi), \mathcal{P}_{m,n}(\tilde{\mathbf{F}})\rangle|=|\langle \mathbf{V}(\theta,\phi), \tilde{\mathbf{F}}\rangle|$, it follows that the beamforming gain at the RX remains the same for any circulant shift applied at the TX. Second, radios at different angular coordinates $(\theta, \phi)$'s, equivalently  different 2D-DFT grid locations $(i,j)$'s, observe different phase changes when circulantly shifting the transmit beamformer. Therefore, as long as the eavesdropper is not in the LoS path between the TX and the RX, the phase change induced at the RX and the eavesdropper are different when circulantly shifting the beamformer. Third, we notice that $\Ntx^2$ distinct 2D-circulant shifts can be applied at the TX for every standard beamformer $\tilde{\mathbf{F}}$. As different circulant shifts induce different phase changes in any direction, our CSB-based defense can randomize the phase at the eavesdropper by applying a random circulant shift of $\tilde{\mathbf{F}}$. It is important to note that circulantly shifting a beamformer at random also induces random phase changes at the RX which is undesirable. 
	
	\par Our CSB-based defense technique determines the phase change induced at the RX apriori, and adjusts the phase of the transmitted symbol accordingly. Such an approach ensures that the RX receives the correct transmitted symbol while the eavesdropper observes a phase perturbed symbol. We define 
	$x'_t = x_t \exp\left(\imj \frac{2\pi}{\Ntx}(m\iMU[t]+n\jMU[t])\right)$
	as the phase adjusted transmit symbol. The symbol $x'_t$ is sent over the beamformer $\mathcal{P}_{m,n}(\tilde{\mathbf{F}}_{t})$ to the RX at 2D-DFT grid location $(\iMU[t],\jMU[t])$. The signal received by the RX can be simplified using \lemmaref{lemma:lemma1} as
	\begin{align}
		y_{\MU,t} & = \sqrt{P_{\rMU[t]}} e^{\imj \nu_\MU}\frob{\mathbf{V}(\thetaMU[t],\phiMU[t]) }{ \mathcal{P}_{m,n}(\tilde{\mathbf{F}}_{t})} x'_t + n_t,  \\
		& = \sqrt{P_{\rMU[t]}} e^{\imj \nu_\MU}\frob{\mathbf{V}(\thetaMU[t],\phiMU[t]) }{ \tilde{\mathbf{F}}_{t}} x_t + n_t.
	\end{align}
	Therefore, by using the circularly shifted beamformer $ \mathcal{P}_{m,n}(\tilde{\mathbf{F}}_{t}) $ and the phase rotated symbol $x'_t$, the received signal at the RX remains unchanged. 
	
\begin{figure}[tb]
		\centering
		\includegraphics[height=2.2in,trim = 50 50 50 50, clip] {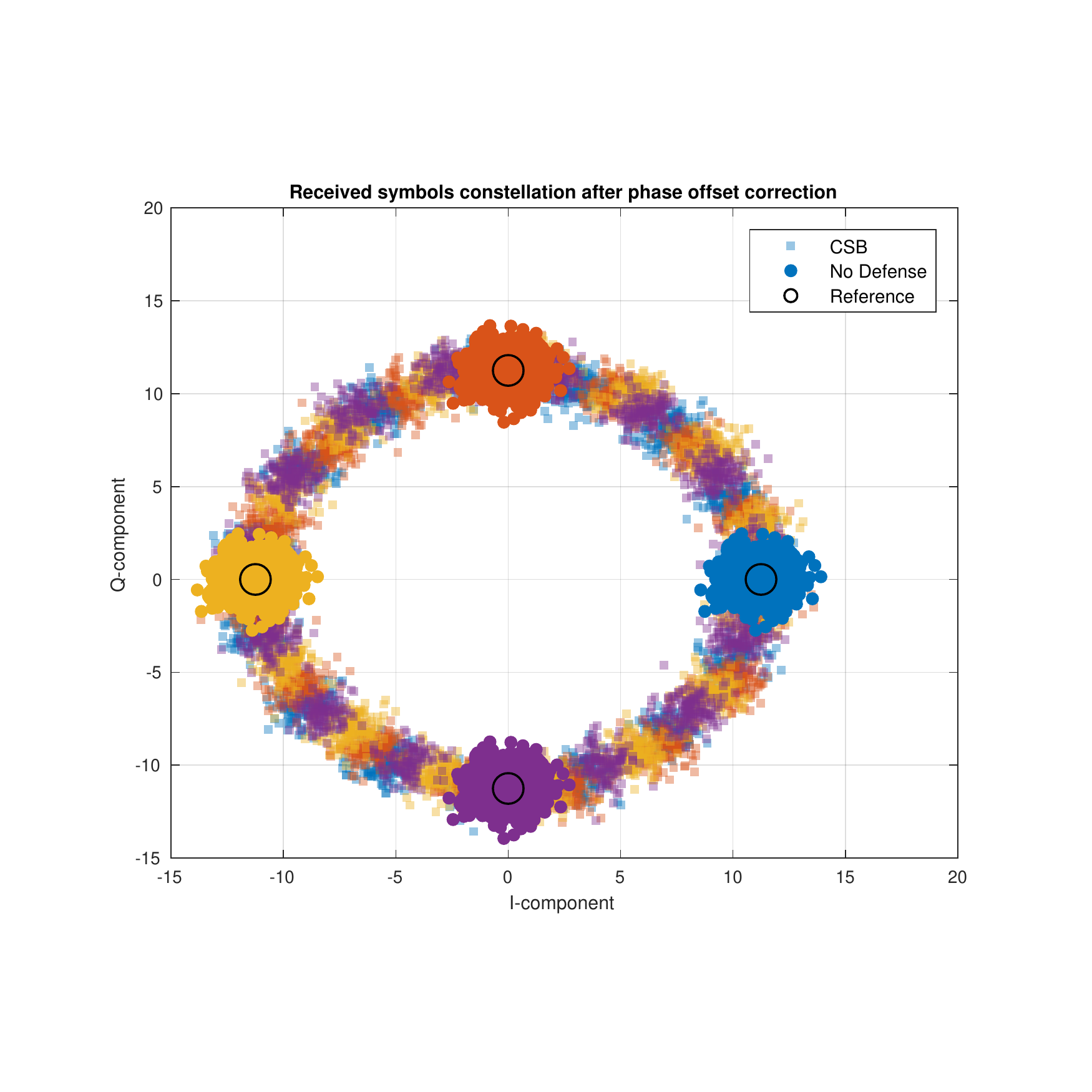}
		\caption{Constellation at the eavesdropper in the presence of \apn induced by CSB, compared to the case without any defense mechanism. CSB applies random circulant shifts of a beamformer to randomize the phase of the symbol at the eavesdropper.}
		\label{fig:constellation}
	\end{figure}

	We now show that CSB perturbs the phase of the symbol received along the directions different from that of the RX. We assume an on-grid eavesdropper and use $(\iEve[t],\jEve[t])$ to denote its 2D-DFT grid location. With the circularly shifted beamformer and the phase-adjusted transmitted symbol, the signal received by the eavesdropper is
	\begin{align}
		y^{(m,n)}_{\Eve,t} & = \sqrt{P_{\rEve[t]}} e^{\imj \nu_\Eve} \frob{\mathbf{V}(\thetaEve[t],\phiEve[t])}{ \mathcal{P}_{m,n}(\tilde{\mathbf{F}}_{t})} x'_t + n_t, \\
		y^{(m,n)}_{\Eve,t} & = \sqrt{P_{\rEve[t]}} e^{\imj \nu_\Eve} \frob{\mathbf{V}(\thetaEve[t],\phiEve[t])}{ \tilde{\mathbf{F}}_{t}} 
		\times x_t e^{\imj \frac{2\pi}{\Ntx} (m (\jMU[t]-\jEve[t]) + n (\iMU[t]-\iEve[t]))} + n_t. \label{eq:eveChange}
	\end{align}
	As the eavesdropper and the RX are located along different directions, we have $(\iMU[t],\jMU[t]) \neq (\iEve[t],\jEve[t])$ for any $t$. In this case, we observe from \eqref{eq:eveChange} that the phase of the symbol received by the eavesdropper is random when the 2D-circulant shift $(m,n)$ is chosen at random. Due to uncertainty in the applied 2D-circulant shift, the eavesdropper cannot predict the induced phase error even with the perfect information of the underlying 2D-DFT beamformer $\tilde{\mathbf{F}}_t$ and the position of the RX $(\thetaMU[t],\phiMU[t])$.
	Therefore, by randomizing the 2D-circulant shifts $(m,n)$ at every symbol and appropriately adjusting the phase of the transmitted symbol, the received signal at the RX is preserved while the phase of the symbol at the eavesdropper is corrupted. An example of the received constellation at the eavesdropper with the CSB technique is shown in \figref{fig:constellation}.
	
	\subsection{Achievable secrecy mutual information}\label{sec:defenseAnalysis}
	In this section, we first characterize the phase errors induced at the eavesdropper and then calculate the \smi achieved by CSB.
	
	We call the phase errors induced by CSB as \apn. We define $\Delta i_t = \iMU[t]-\iEve[t]$ and $\Delta j = \jMU[t] - \jEve[t] $ as the difference in the DFT grid coordinates corresponding to the RX and the eavesdropper. The error in the phase of the received symbols at the eavesdropper, i.e., the \apn, can be expressed using \eqref{eq:eveChange} as
	\begin{equation}
		\Delta \Phi_t = \frac{2\pi}{\Ntx} \modulo{ m \Delta i_t + n\Delta j_t}.
	\end{equation}
	We also define $g_t = \gcd(\Delta i_t, \Delta j_t)$. In Lemma \ref{lemma:lemma2}, we derive statistical properties of the \apn. 
	We avoid the subscript $t$ for simplicity of notation.
	\begin{lemma}\label{lemma:lemma2}
		Consider independent random variables $M_0$ and $N_0$ that are uniformly distributed over $\Omega = \set{\Ntx}$. We define $g = \gcd(\Delta i, \Delta j)$,
		\begin{equation}
			\Delta \Phi = \frac{2\pi}{\Ntx} \modulo{ M_\mathrm{0} \Delta i + N_\mathrm{0}\Delta j}, \, \, \mathrm{and}\, \,  
			\Omega_{\Phi_{g}} = \left\{\frac{2\pi\modulo{g i}}{\Ntx}  : \forall i \in \set{\frac{\Ntx}{\gcd(\Ntx,g)}} \right\}.
		\end{equation}
		Then,
		\begin{equation}
			\mathbb{P} \left(\Delta \Phi = \phi \right) = \begin{cases}
				\frac{\gcd(\Ntx,g)}{\Ntx}, & \phi\in \Omega_{\Phi_{g}}\\
				0, & \mathrm{otherwise}
			\end{cases}.
		\end{equation}
		\begin{proof}
			The proof contains two steps: $(\mathsf{i})$ For any pair $(m,n)\in\set{\Ntx}^2,\, \Delta\Phi \in \Omega_{\Phi_{g}}.$ 
			$(\mathsf{ii})$ If the random variables $M_0, N_0$ are uniformly distributed, then $\Delta \Phi$ is uniformly distributed over $\Omega_{\Phi_{g}}$. 
			
			\par We prove the first step $(\mathsf{i})$ by induction. For the case $(m,n) = (0,0)$, $\Delta\Phi = 0 \in \Omega_{\Phi_{g}}$. We assume that for the pair $(m,n)$, $\Delta\Phi = \frac{2\pi}{\Ntx} \modulo{m\Delta i + n\Delta j} = \frac{2\pi\modulo{g \ell}}{\Ntx}$, where $\ell$ is some integer. Then, for the pair $(m+1,n)$,
			\begin{align}
				\Delta\Phi' & = \frac{2\pi}{\Ntx} \modulo{(m+1)\Delta i + n\Delta j}\\
				& = \frac{2\pi}{\Ntx} \modulo{\modulo{m\Delta i + n\Delta j}+\modulo{\Delta i}}\\
				& \overset{(a)}{=} \frac{2\pi}{\Ntx} \modulo{\modulo{g \ell}+\modulo{g k}} = \frac{2\pi}{\Ntx} \modulo{g (\ell +k)} \in \Omega_{\Phi_{g}},
			\end{align}
			where the equality $(a)$ uses the fact that $\Delta i = g k$ for some integer $k$ if $g = \gcd(\Delta i, \Delta j)$. 
			Therefore, if there exists a pair $(m,n)$ such that $\Delta\Phi \in \Omega_{\Phi_{g}}$, $\Delta\Phi'$ corresponding to the pair $(m+1,n)$ belongs to $\Omega_{\Phi_{g}}$. Similarly, it can be shown that $\Delta\Phi'$ corresponding to $(m,n+1)$ also belongs to $\Omega_{\Phi_{g}}$. Therefore, it follows by induction that $\Delta\Phi \in \Omega_{\Phi_{g}}$ for every $(m,n)\in\set{\Ntx}^2$.
			
			\par We now prove the second step $(\mathsf{ii})$ in Lemma \ref{lemma:lemma2}. To show that $\Delta \Phi$ is uniformly distributed over $\Omega_{\Phi_{g}}$, we prove that there are same number of $(m,n)$ pairs such that $\Delta \Phi = \frac{2\pi \modulo{g\ell}}{\Ntx}$ for any $\ell$. 
			We denote by $m_0, n_0$ as the smallest values of $m, n$ that satisfy $\modulo{m\Delta i + n\Delta j}=\modulo{g \ell}$, i.e., $m_0 \Delta i + n_0 \Delta j = g\ell + k\Ntx$, for some integer $k\geq 0$. We also consider an integer pair $(k_1, k_2)$, such that $(\mathsf{i}) \, \frac{k_1 \Ntx}{\Delta i}, \frac{k_2\Ntx}{\Delta j} \leq \Ntx-1$, $(\mathsf{ii}) \, \frac{k_1 \Ntx}{\Delta i}$, $\frac{k_2\Ntx}{\Delta j}$ are integers, and $(\mathsf{iii}) \, k_1/\Delta i + k_2/\Delta j$ is an integer. Then, \begin{align}
				{\left(m_0+k_1\frac{\Ntx}{\Delta i}\right) \Delta i +\left(n_0+k_2\frac{\Ntx}{\Delta j}\right)\Delta j} & = g\ell + (k+r)\Ntx,
			\end{align}
			where $r$ is some integer. Thus, for each permissible pair $(k_1,k_2)$, there exists a pair $(m,n) = (m_0 + \frac{k_1 \Ntx}{\Delta i}, n_0 + \frac{k_2\Ntx}{\Delta j})$ such that $\Delta\Phi = \frac{2\pi \modulo{g\ell}}{\Ntx}$. 
			Observe that the number of permissible pairs $(k_1, k_2)$ only depend on $\Delta i, \Delta j, \Ntx$, and \textit{not} on $\ell$. 
			Therefore, for every $\ell$, there are same number of $(m,n)$ pairs, such that $\Delta\Phi = \frac{2\pi \modulo{g\ell}}{\Ntx}$. 
			As a result, by choosing the pair $(m,n)$ uniformly from $\set{\Ntx}^2$, it can be ensured that $\Delta \Phi$ is uniformly distributed over $\Omega_{\Phi_{g}}$.%
		\end{proof}
	\end{lemma}
	
	\lemmaref{lemma:lemma2} shows that the \apn induced by CSB is uniformly distributed over $\Omega_{\Phi_{g}}$. With this result, we show in \lemmaref{lemma:lemma3} that the \apn introduced by CSB renders the eavesdropper unable to distinguish the transmitted symbol from the phase-corrupted received symbol. 
	
	\begin{lemma}\label{lemma:lemma3}
		Consider an $M$-PSK constellation with the symbol set $\mathcal{M}$. We  define partitions of $\mathcal{M}$ such that each partition contains $\gcd\left(\left\vert\Omega_{\Phi_{g}}\right\vert,M\right)$ number of symbols spaced uniformly in phase. The eavesdropper cannot distinguish between the symbols within a partition due to the \apn induced by CSB. Additionally, there are $M/\gcd(|\Omega_{\Phi_g}|,M)$ number of symbols that can be accurately distinguished. 
		\begin{proof}
			To prove this lemma, we first find a condition when two symbols $e^{\imj 2\pi k_1/M}$ and $e^{\imj 2\pi k_2/M}$ in a constellation $\mathcal{M}$ cannot be distinguished due to the \apn induced by CSB. For two symbols to be indistinguishable under \apn, the difference in the phases of the both symbols must be in $\Omega_{\Phi_g}$. Equivalently, 
			\begin{equation}
				\frac{2\pi k_1}{M} - \frac{2\pi k_2}{M} = \frac{2\pi\modulo{g \ell}}{\Ntx} + 2\pi p_1,
			\end{equation}
			where $p_1$ is an integer and $\ell \in \set{\frac{\Ntx}{\gcd(\Ntx, g)}}$. Observe that $\modulo{g\ell} + p_2 \Ntx = g\ell$, for some integer $p_2$. As a result, we can write
			\begin{align}
				\frac{k_1 - k_2}{M} - \frac{g\ell}{\Ntx} & = p_1 - p_2 := p_3.\label{eq:lemma3_1}
			\end{align}
			We define $g' = \gcd(g,\Ntx)$. Then $\Ntx = g' u_1$ and $g = g' u_2$, for some integers $u_1, u_2$. Additionally, note that $u_1 = |\Omega_{\Phi_g}|.$
			By re-arranging \eqref{eq:lemma3_1}, we get
			\begin{align}
				\frac{|\Omega_{\Phi_g}|}{M} (k_1-k_2) - u_2 \ell = |\Omega_{\Phi_g}| p_3.\label{eq:lemma3_2}
			\end{align}
			To satisfy \eqref{eq:lemma3_2}, $(k_1 - k_2)$ must be an integer multiple of $M/\gcd(M,|\Omega_{\Phi_g}|)$. 
			We define a partition of constellation $\mathcal{M}$, denoted by $\mathcal{M}_{k_1}$ containing the symbol $e^{\imj\frac{2\pi k_1}{M}}$, and all symbols $e^{\imj\frac{2\pi k_2}{M}}$ such that $k_1 - k_2$ satisfies \eqref{eq:lemma3_2}. Specifically, 
			\begin{equation}
				\mathcal{M}_{k_1} = \left\{\exp\left(\imj\frac{2\pi k_1}{M} +\imj \frac{2\pi i}{\gcd(M,|\Omega_{\Phi_g}|)}\right): i\in \set{\gcd(M,|\Omega_{\Phi_g}|)}\right\}.
			\end{equation}
			Note that each partition contains $\gcd(M,|\Omega_{\Phi_g}|)$ number of symbols that cannot be distinguished from other symbols in that partition. 
			Furthermore, there are $M/\gcd(M,|\Omega_{\Phi_g}|)$ number of partitions. As a result, out of the $M$ symbols in the constellation $\mathcal{M}$, $M/\gcd(M,|\Omega_{\Phi_g}|)$ number of symbols are distinguishable under \apn. We explain the interpretation of this lemma using \exref{ex:ex1}.  
		\end{proof}
	\end{lemma}
	\begin{example}\label{ex:ex1}
		Consider a TX with $\Ntx = 16$ that uses a QPSK constellation. In the high \snr regime at the eavesdropper, the mutual information transfer to the eavesdropper is $\log_2(4/\gcd(|\Omega_{\Phi_{g_t}}|, 4))$ bits/symbol. 
		If $g_t \notin \{0,8\}$, the mutual information between the TX and the eavesdropper is $0$ bit/symbol. Alternatively, if $g_t = 8$ the mutual information between the TX and the eavesdropper is $1$ bit/symbol. 
		Therefore, with CSB defense, the eavesdropper can only receive meaningful information along the certain directions associated with $g_t = 8$ and $g_t = 0$. Combined with directional beam patterns, the performance of eavesdropper is limited by low energy leakage or high phase corruption. 
	\end{example}
	
	We now use \lemmaref{lemma:lemma3} to derive the \smi with CSB defense by considering an $M$-PSK constellation. 
	The \smi, measured in bits/symbol, is defined as the difference between the information transferred over the TX-RX channel and the TX-eavesdropper channel. We denote mutual information (MI) of the channel between TX and RX by $\mathcal{I}_\MU$, and MI of the channel between TX and eavesdropper by $\mathcal{I}_\Eve$. Thus, we can define the \smi ${C}_\mathrm{S}$ at time $t$ as
	\begin{equation}\label{eq:firstSecrecyRateSeq}
		C_\mathrm{S}(t) = \max\left\{\mathcal{I}_\MU - \mathcal{I}_\Eve, 0\right\}
	\end{equation}
	We define $\mathcal{I}(\rho, M)$, measured in bits per symbol, as the spectral efficiency of the channel with \snr $\rho$ and the input $M$-PSK constellation~\cite{Hea:Introduction-to-Wireless-Digital:17}. 
	Additionally, if the eavesdropper is located at an on-grid position at time $t$ such that $ \gcd(\Delta i_t, \Delta j_t) = g_t $, then from \lemmaref{lemma:lemma3}, communication over the CSB-secured TX-eavesdropper channel using $M$-PSK modulation is equivalent to communication over the unsecured TX-eavesdropper channel using $M/\gcd(g_t, M)$-PSK constellation. Thus, if the angular coordinate of the RX at time $t$ is $(\thetaMU[t], \phiMU[t])$, and that of the eavesdropper is $(\thetaEve[t],\phiEve[t])$, then using beamformer $\tilde{\mathbf{F}}_t$ at time $t$, we can calculate the \smi with CSB defense as
	\begin{align}\label{eq:finalSecrecyRateSeq}
		C_\mathrm{S}(t) = \max&\left\{\mathcal{I}\left(\frac{P_{\rMU[t]}}{\sigma^2}\left\vert\frob{\mathbf{V}(\thetaMU[t],\phiMU[t]) }{\tilde{\mathbf{F}}_t }\right\vert^2, M\right)\right. \nonumber\\ 
		&\left.-\mathcal{I}\left(\frac{P_{\rEve[t]}}{\sigma^2}\left\vert\frob{\mathbf{V}(\thetaEve[t],\phiEve[t]) }{\tilde{\mathbf{F}}_t }\right\vert^2,\frac{M}{ \gcd \left(\left\vert \Omega_{\Phi_{g_t}} \right\vert, M\right)}\right),0\right\}.
	\end{align}
	For an effective eavesdropping attack, the eavesdropper attempts to minimize $C_\mathrm{S}(t)$ by positioning itself to appropriate $(\thetaEve[t],\phiEve[t])$. In the presence of CSB defense, the position of the eavesdropper, however, affects not only the \snr at the eavesdropper but also $|\Omega_{\Phi_{g_t}}|$, i.e., the equivalent constellation observed by the eavesdropper. Thus, CSB defense reduces information transfer to the eavesdropper by corrupting the constellation. 
	
	\noindent\textbf{Remark:} For the design of the CSB defense, we considered a narrowband single-path channel. In a multi-path environment with different angle of departures, the RX receives a combination of desired constellation and a phase perturbed constellation. Due to the use of directional beams at the TX, however, the signals received from the non-dominant paths will have significantly less energy, thereby resulting in small perturbations in the constellation at the RX. 
	
	\subsection{Implementing CSB - A packet level overview}
	\begin{figure}[tb]
		\centering
		\includegraphics[height=0.85in,width=0.85\columnwidth,keepaspectratio]{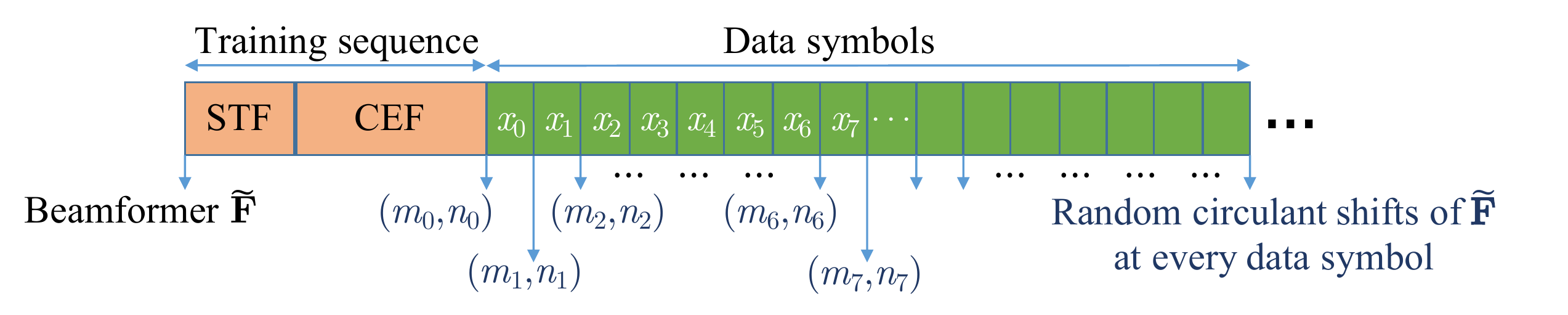}
		\caption{IEEE 802.11ad packet structure: CSB defense uses $(m,n)$-2D circulantly shifted beamformers, where $(m,n)$ are IID random variables from the set $\set{\Ntx}^2$. Circulantly shifting a beamformer at every data symbol distorts the constellation at the eavesdropper, even with perfect CFO correction and channel estimation.}
		\label{fig:implementationDetails}
	\end{figure}
	In this part, we describe the details related to implementation of CSB. \figref{fig:implementationDetails} describes a typical PHY layer packet structure in IEEE 802.11ad protocol~\cite{NitCorFlo:IEEE-802.11ad:-directional:14}. The training sequences, mainly short training field (STF) and channel estimation field (CEF), are used for the frame synchronization, carrier frequency offset (CFO) and phase offset correction. Then, data symbols are transmitted by the TX, followed by another packet or a short beam training field. 
	
	We propose to use CSB defense during the data symbol transmission. Specifically, the TX uses a fixed beamformer $\tilde{\mathbf{F}}$ for transmission of the training sequence. It allows the RX to perform frame synchronization, CFO and phase offset corrections, and channel estimation. Then, during data transmission, the TX circulantly shifts the beamformer by $(m,n)$ units. Here, $(m,n)$ is chosen at random from the set $\set{\Ntx}^2$ for each data symbol. For a particular $(m,n)$ shift, the phase of the transmitted symbol is adjusted such that the phase of the symbol received in the direction of the RX remains unchanged. Thus, the RX receives the data symbols in a way that is agnostic to the circulant shifts applied at the TX. The eavesdropper, however, suffers from phase errors induced due to circulant shifting. Although using a fixed beamformer to transmit the training sequence allows the eavesdropper to equalize the channel, the symbols received by the eavesdropper are distorted due to circulant shifting of the beamformer. 
	
	In case of an OFDM-based operation with 802.11ad, CSB introduces the same phase error across all the sub-carriers as analog beams are frequency flat. Under a constant phase perturbation, the eavesdropper can correct the phase of the received OFDM symbol using pilot sub-carriers. To overcome this attack, the TX can leverage the large symbol period of an OFDM symbol to circulantly shift the beamformer multiple times within a symbol period. By adjusting the phase of the transmitted symbol after every shift of the beamformer, the received OFDM symbol is corrupted along all directions other than the direction of the RX. 
	
	\section{Experimental Validation}\label{sec:experiment}
	In this section, we design an experiment to validate the premise of CSB defense. Specifically, our experiment estimates the phase change induced by circularly shifting a beamformer and shows that the estimated phase change is consistent with the result in \lemmaref{lemma:lemma1}.
	
	\subsection{Hardware setup}
	We use two N210 USRPs, each as the baseband processor at the transmitter and the receiver. Each USRP is connected to a separate SiBEAM Sil6342 phased array operating at 60.48 GHz. These phased arrays are uniform linear arrays with 12-antenna elements. Each element is connected to a 2-bit phase shifter that can be configured independent of the others. 
	\begin{figure}[tb]
		\centering
		\includegraphics[height=2in]{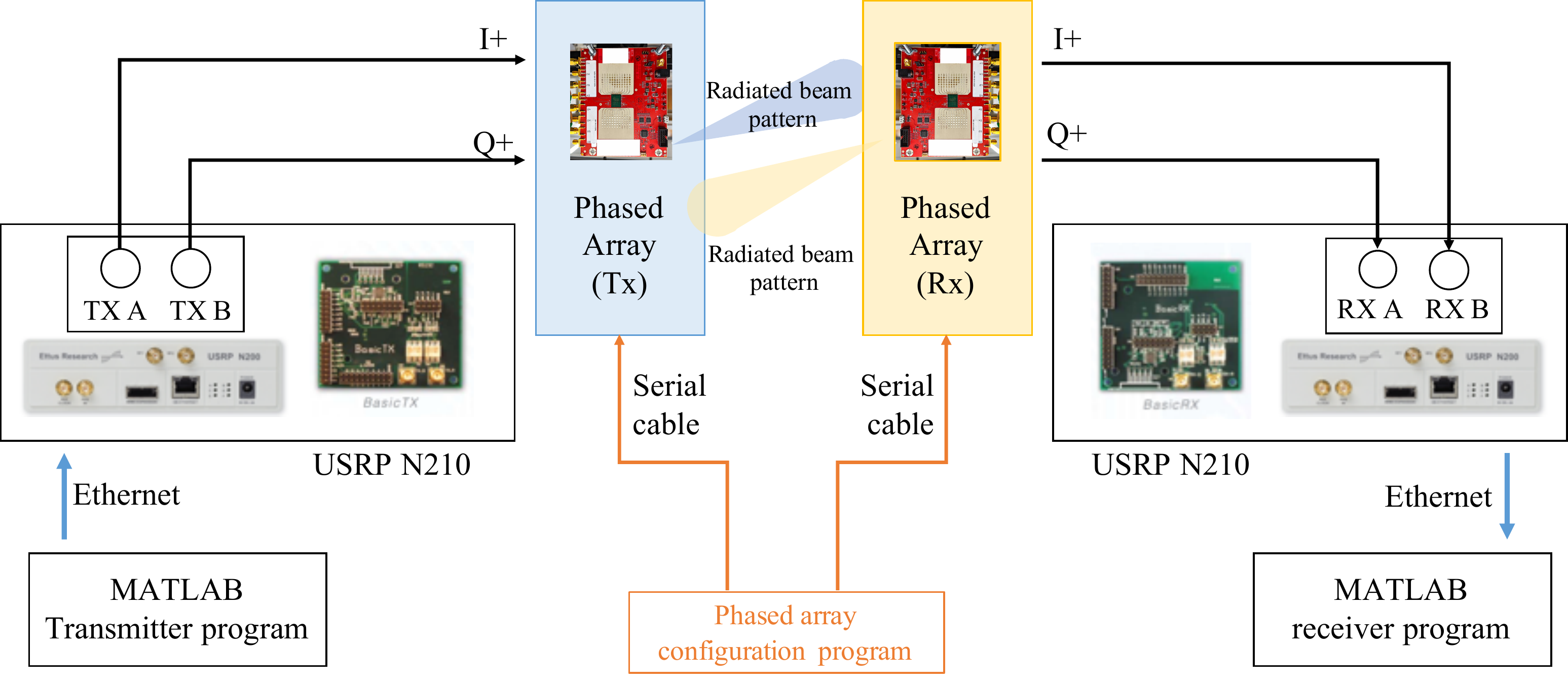}
		\caption{Block diagram of the experimental setup used to validate the key idea in CSB defense.}
		\label{fig:HardwareSetup}
	\end{figure}
	A block diagram with the hardware connections is shown in \figref{fig:HardwareSetup}. We use the following procedure to setup the transmitter: 
	$(\mathsf{i})$ A MATLAB instance runs the transmitter program and generates the I/Q samples that are sent to USRP via Ethernet cable. $(\mathsf{ii})$ The USRP then generates the baseband signal that is fed into the transmitter phased array.
	$(\mathsf{iii})$ The phased array configuration program (external to the transmitter program) sets the configuration of the phase shifters using a universal asynchronous receiver-transmitter (UART) protocol.
	$(\mathsf{iv})$ The baseband signal is upconverted to 60.48 GHz and the upconverted signal is phase shifted with the set configuration of the phased array. Finally, the $12 \times 1$ phase shifted signals are transmitted over the channel.
	A similar setup $(\mathsf{i})-(\mathsf{iv})$ is built at the receiver.
	
	The SiBEAM Sil6342 phased arrays allow reconfiguration of the phase shifters using a UART protocol. The phase shifter of each antenna element can be set to one of the four phase states. The combination of the phase states applied to the $12\times 1$ phased array realizes a specific beamformer. For the experiment, we emulate a one-bit phased array by using only two states out of four available phase states. Using one-bit phased array allows us to analytically predict the leaked RF signal which is mirror symmetric to the target direction as proven in Appendix \ref{appendix:onebitreflection}. Unlike ideal phased arrays, the off-the-shelf phased array used in our experiment does not provide the precise phase shifts of $\{0,\pi \}$ due to hardware imperfections. The phase offsets from $0$ and $\pi$ are estimated at each antenna using the calibration procedure described in \cite{ZhaPatSha:Side-Information-Aided-Noncoherent-Beam:19}. 
	With the knowledge of the phase offsets associated with the phase states, the phase of every entry in the beamformer is mapped to the nearest phase offset available at that antenna element.

	\subsection{Experimental procedure}
	\begin{figure}[tb]
		\centering
		\subfigure[\label{fig:ExperimentMeasurementPacket}]{
			\includegraphics[height=1.5in]{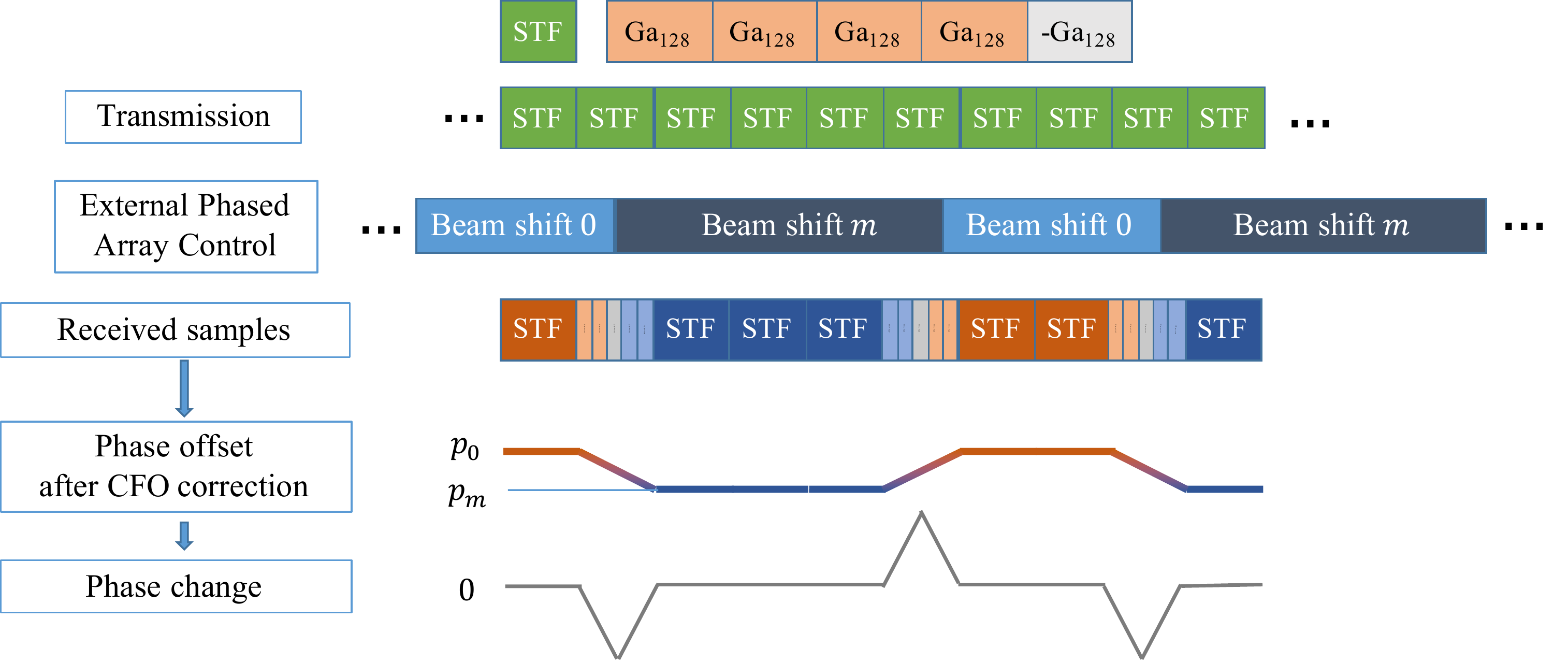}
		}\iftwocolumn{\\}
		\subfigure[\label{fig:ExperimentPhaseChange}]{
			\includegraphics[height=1.5in]{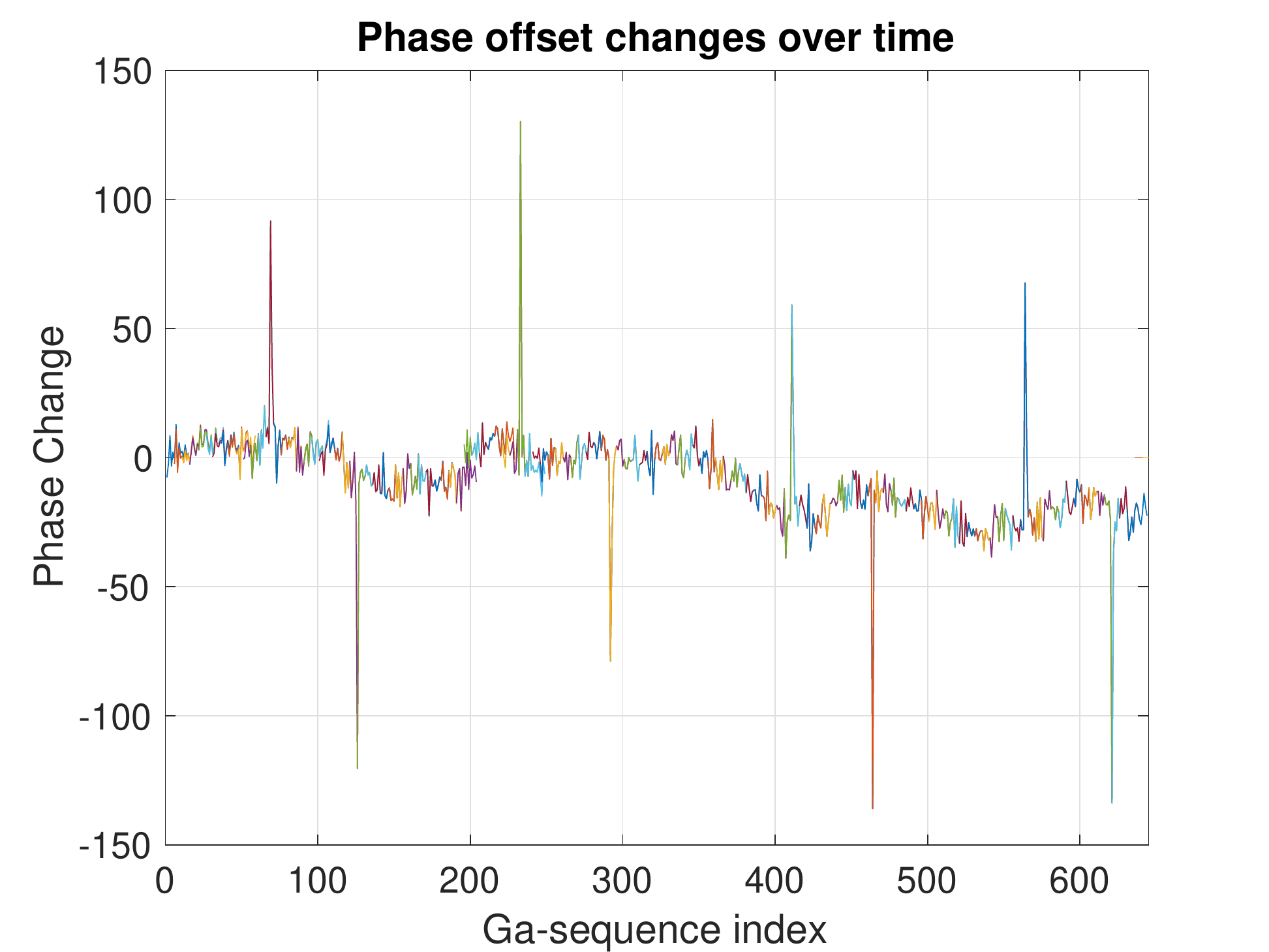}
		}
		\caption{(a) The figure describes our experimental procedure in which the phased arrays are externally controlled. It also shows the phase offset induced due to change in the beamformer. (b) This plot shows the phase difference between consecutive Ga-sequences in a packet. The spikes indicates changes in the beamformer and second, fourth, and sixth spike indicates transition from the test beamformer to its $m$-circulant shift.}
	\end{figure}
	
	In \figref{fig:ExperimentMeasurementPacket}, we describe the packet structure and the experimental procedure. A packet consists of a group of 130 short training fields (STF) where each STF contains five 128-length Golay sequences. The TX transmits an uninterrupted stream of identical packets while the receiver captures one packet at a time. To accurately measure the phase change due to circulant shifting of the beamformer, it is important to maintain coherence across measurements acquired before and after the circulant shift. Any interruption during packet reception must be avoided as it introduces a phase noise that cannot be corrected. To this end, we design our experiment by separating the receive operation and the circulant shifting of the transmit beamformer. Specifically, the receiver acquires a packet without any interruption, while an external program periodically applies beamformers by alternating between a beamformer and its $m$-circulant shift within the same packet. 
	
	To estimate the phase change due to the change in the beamformer, we first correct the frequency offset of each STF, and calculate the phase offset of each Ga-sequence in an STF. As a result, any significant change in the phase offsets of consecutive Ga-sequences can be attributed to circulantly shifting the beamformer. 
	The measured phase change is either due to $(\mathsf{i})$ the transition from the test beamformer to its $m$-circularly shifted beamformer or  $(\mathsf{ii})$ the transition from the $m$-circularly shifted beamformer to the test beamformer. 
	To distinguish between the two phase changes, we use different dwell durations for the test beamformer and its $m$-circulant shift. In particular, we implement the test beamformer for $1/3$rd of the period duration and its $m$-circulant shift for $2/3$rd of the period duration. Under such a setting, if two consecutive phase changes occur at a lag of $1/3$rd of the period duration, we can conclude that the later phase change is due to the transition from the test beamformer to its $m$-circulant shift. 
	
	In \figref{fig:ExperimentPhaseChange}, we show the difference between the phase offsets of consecutive Ga-sequences within a packet. The periodic pairs of spikes indicate sudden changes in the phase offset of consecutive Ga-sequences. These jumps are due to change in the beamformer. Furthermore, the long duration after the second, fourth and sixth spike is due to the transition from the beamformer to its $m$-circulant shift. By measuring the changes in the phase offsets and averaging them, we get the phase change due to the transition from the test beamformer to its $m$-circulant shift along a direction. 
	Similarly, we measure the phase shift along different directions for every $m\in\{1,2,...,11\}$.
	
	\subsection{Experimental results}
	\begin{figure}[tb]
		\centering
		\includegraphics[height=1.8in,width=0.7\columnwidth,keepaspectratio]{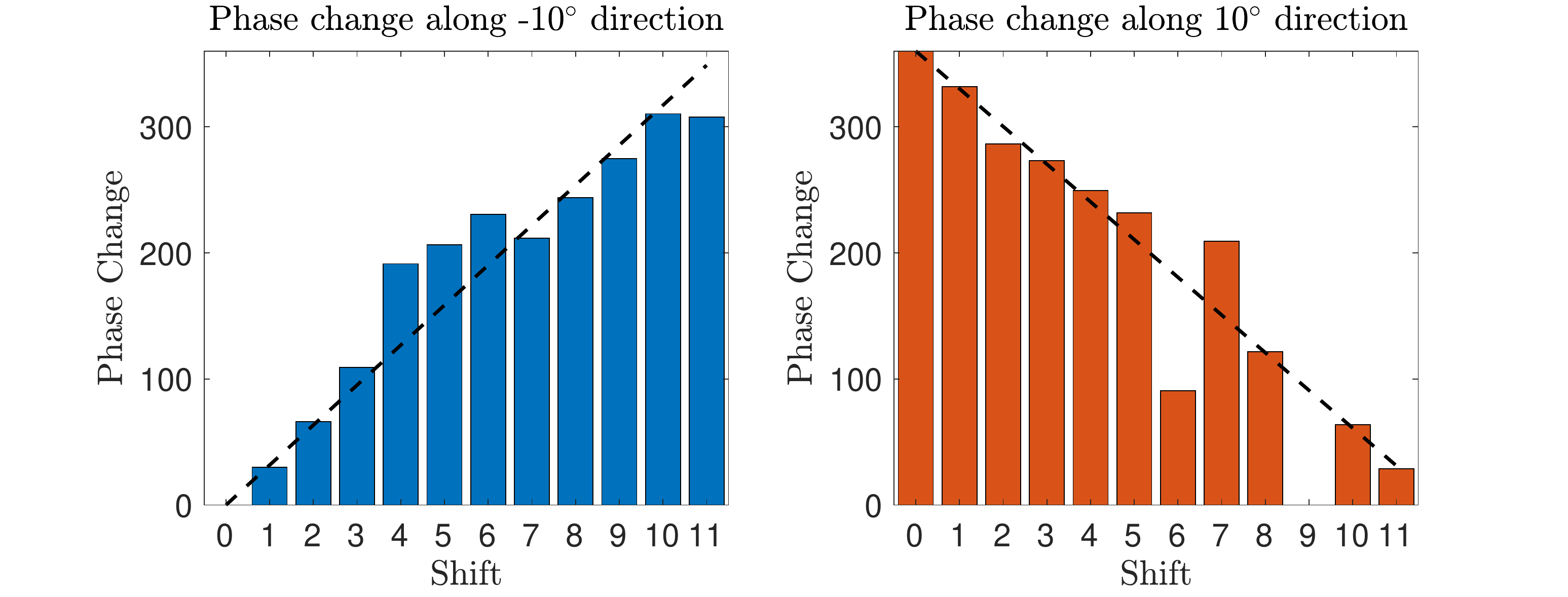}
		\caption{Phase change corresponding to different circulant shifts in the beamformer when the receiver is at $10^\circ$ and $-10^\circ$ with respect to the boresight of the transmit phased array. The estimated slopes of the dotted lines are $31.7^\circ$ and $-29.9^\circ$ per shift. These estimates are close to their theoretical values of $30^\circ$ and $-30^\circ$ per shift.}
		\label{fig:ExperimentMainResult}
	\end{figure}
	
	The measurements collected using our experimental procedure are post-processed to verify \lemmaref{lemma:lemma1}. For the experiment, we use a one-bit quantized beamformer ($q=1$) for directional beamforming along $10^\circ$ relative to the boresight. Due to the one-bit quantization, the beam pattern is symmetric about the boresight, i.e., the beam has two main lobes at $10^\circ$ and $-10^\circ$. Different circulant shifts of this beamformer are applied at the TX. In each case, the phase change induced due to circulant shift is measured by placing a receiver at $10^\circ$. The experiment is repeated again by moving the receiver to $-10^\circ$. From \figref{fig:ExperimentMainResult}, we observe that the phase change is linear with applied circulant shift $m$ as derived in \lemmaref{lemma:lemma1}. The slope of this linear variation is also consistent with the angle from the boresight, as shown in \figref{fig:ExperimentMainResult}. As the phase change induced at the RX by circulantly shifting a transmit beamformer can be predicted, the phase of the transmitted symbols can be adjusted at the TX for correct decoding along the direction of the RX. Such an adjustment, however, does not correct the phase perturbation at the eavesdropper. This is because the phase change induced by circulant shifting a beamformer is different along different directions.
	
	
	\section{AirSpy: An Attack on V2I network}\label{sec:attack}
	In this section, we describe an attack, called \textit{AirSpy}, on a planar low-resolution phased array TX in a downlink V2I network. We assume a mobile UAV eavesdropper that is aware of the resolution of the RF phased array at the TX and the position of the RX. The attack is achieved by computing a UAV flight path that efficiently taps the leaked RF signals in a mechanically feasible manner. 
	%
	We first define the secrecy rate of the link between the TX and the RX. Then, we develop an attack by formulating a trajectory search problem under the mechanical constraints on the UAV. Finally, we discuss a dynamic programming-based algorithm for trajectory search. 
	
	\subsection{Secrecy rate}
	To measure the severity of a physical layer attack, we define the secrecy rate corresponding to a beamformer $\tilde{\mathbf{F}}_t$ as 
	\begin{align}
		\mathcal{C}(\tilde{\mathbf{F}}_t, (\rEve[t],\thetaEve[t],\phiEve[t])) %
		& = \left[ \log \left(1+\frac{P_{\rMU[t]}}{\sigma^2} \left\vert\frob{\mathbf{V}(\thetaMU[t],\phiMU[t]) }{\tilde{\mathbf{F}}_t }\right\vert^2\right) \right.\nonumber\\
		& \left. \quad- \log\left(1+\frac{P_{\rEve[t]}}{\sigma^2} \left\vert\frob{\mathbf{V}(\thetaEve[t],\phiEve[t]) }{ \tilde{\mathbf{F}}_t }\right\vert^2\right) \right].\label{eq:secrecyRate}
	\end{align}
	A greedy attack strategy is one that finds an optimal eavesdropping position $(\thetaEve,\phiEve) \neq (\thetaMU[t],\phiMU[t])$ which minimizes the secrecy rate at every time instant. Such a greedy approach, however, may be mechanically infeasible under a finite velocity constraint.  
	A good attack strategy is one that identifies and tracks multiple RF leakage signals over time for long term exploitation under the velocity constraint. 
	


	\subsection{Learning algorithm for eavesdropping trajectory design}
	In this section, we define a trajectory and the set of feasible trajectories that satisfies the mechanical constraints on the motion of the UAV. Then, we propose an efficient dynamic programming-based algorithm that finds a UAV trajectory to eavesdrop on the TX. Our design assumes perfect knowledge of the RX location over a time interval, and minimizes the sum secrecy rate in this interval.
	
	\begin{figure}[tb]
		\centering
		\includegraphics[height=1.8in]{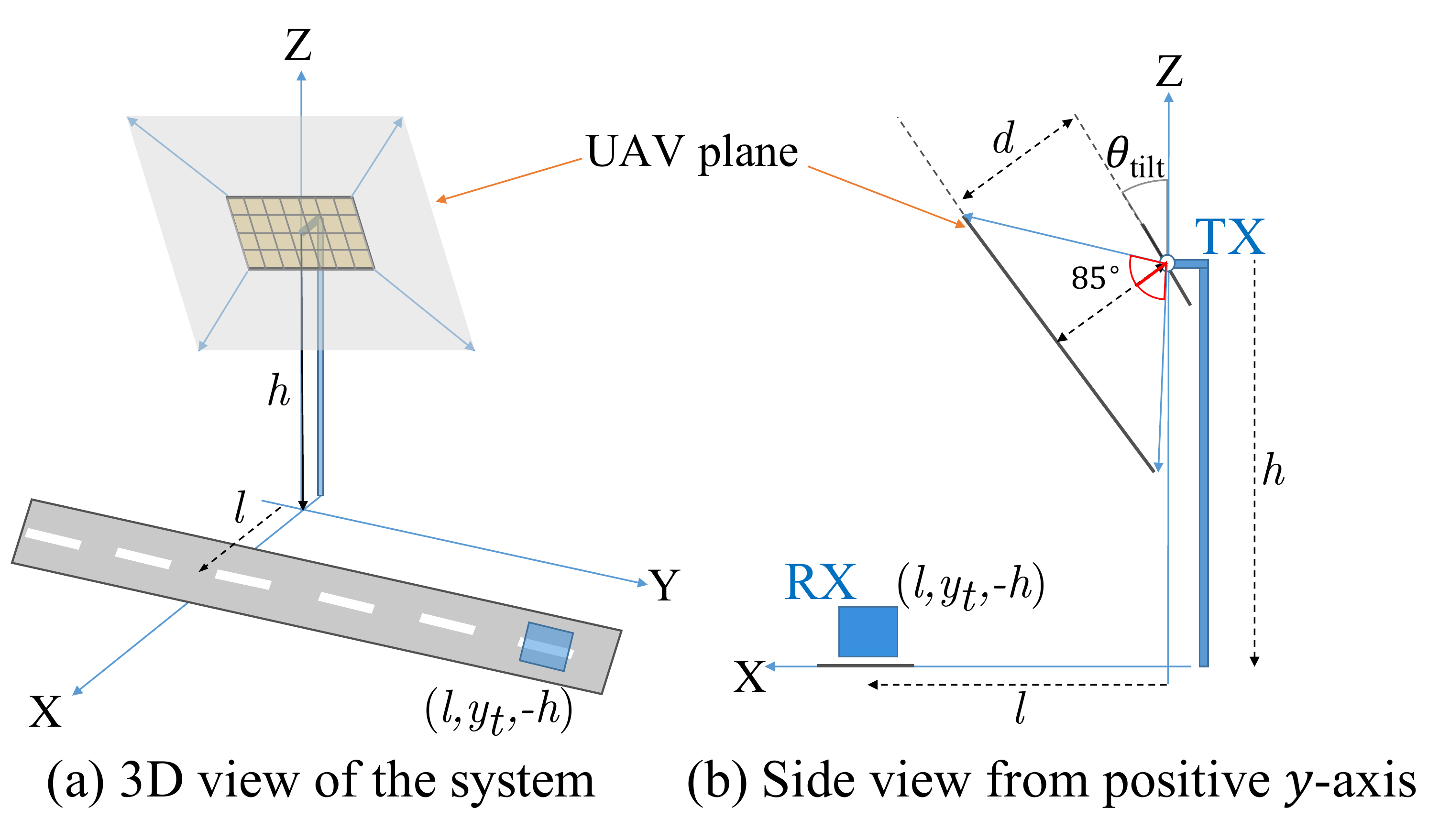}
		\caption{We assume that the UAV moves on a plane parallel to the plane of TX antenna array at a distance $d$. The angle subtended by the UAV plane at the center of the TX antenna array is $\beta/2 = 85^\circ$.}
		\label{fig:UAVPlane}
	\end{figure}
	
	We consider a TX equipped with a planar antenna array situated at a height $h$ from the ground. We assume that the RX is a vehicular receiver that travels on a linear ground trajectory defined by the line $\{x=\ell, z = -h\}$. To incorporate the mechanical constraints on the eavesdropping UAV and design a numerically efficient algorithm, we limit the motion of the UAV to a virtual plane called the \textit{UAV Plane}. This plane is parallel to the plane of the TX antenna array at a distance $d$, as shown in \figref{fig:UAVPlane}. The azimuth and elevation angles subtended by the UAV plane at the center of the TX antenna array are both equal to $\beta$, where $\beta \in (0,\pi)$. We use $\mathcal{P}_d$ to denote the set of points on the UAV plane, i.e., 
	\begin{align}
		\mathcal{P}_d = \left\{\right.&(x,y,z): x\cos \thetatilt - z \sin \thetatilt = d, \nonumber\\
		& \phi = \arctan (z/x) + \thetatilt \in [-\beta/2, \beta/2], \theta = \arctan (y/x) \in \left. [-\beta/2, \beta/2]\right\}
	\end{align}
	For any angular coordinate of the eavesdropper $(\thetaEve, \phiEve)\in[-\beta/2,\beta/2]^2$, there is a unique 2D-coordinate on the UAV plane. With the UAV plane constraint, the eavesdropper trajectory design problem is simplified from 3D to 2D. 
	
	We use a 2D coordinate system centered at the UAV plane to denote points on the UAV plane. The 2D-coordinate $(u,v)$ corresponding to $(x,y,z)\in \mathcal{P}_d$ is computed from $
	x_u  = u d\tan(\beta/2)\allowbreak\sin\thetatilt + d\cos\thetatilt,\ 
	y_v  = v d\tan(\beta/2),\ 
	z_u  = u d\tan(\beta/2)\allowbreak\cos\thetatilt - d\sin\thetatilt,
	$ where $u,v\in[-1,1]$. For notational convenience, we define a mapping $S_2: [-1,1]^2 \rightarrow \mathcal{P}_d$ such that $(x_u,y_v,z_u) = S_2(u,v)$. We discretize the time index $t$ with a sampling period $\Ts$, and minimize the sum secrecy rate over discrete time instances for computational tractability. For that, we define a trajectory in \defref{def:traj}. 
	\begin{definition}\label{def:traj}
		A discrete trajectory of length $N$, denoted by $\tau_{N,d}$, is a sequence of $(u_t,v_t)$ pairs, where $(u_t,v_t) \in [-1,1]^2$ and $t=0,1,\ldots, N-1$, such that $t$-th element of the sequence represents the coordinate of the UAV with respect to the center of the UAV plane at time $t\Ts$. We denote $t$-th element of the trajectory $\tau_{N,d}$ by $\tau_{N,d}(t) = (u_t,v_t)$.
	\end{definition}
	We would like to mention that only a subset of the trajectories in \defref{def:traj} are permissible for the UAV. First, the trajectory must meet the maximum permissible velocity constraint on the UAV. Second, the UAV following this trajectory should not block the LoS path between the TX and the RX at any time instant. Based on these constraints, we define the set of permissible trajectories in \defref{def:permissible_traj}. Recall that the mapping $S_1$ converts rectangular coordinates to modified spherical coordinates, and $S_2$ changes the reference from the center of the UAV plane to the center of the TX antenna array. 
	\begin{definition}\label{def:permissible_traj}
		Let $\vmax$ be the maximum permissible velocity of the UAV, $(\thetaMU[t],\phiMU[t])$ be the angular coordinate of the RX with respect to TX at time $t$, and $(r_t,\theta_t,\phi_t)$ denote the angular coordinate of the UAV such that $(r_t,\theta_t,\phi_t) = S_1(S_2(u_t,v_t))$. Then, a discrete trajectory $\tau_{N,d}$ is a \textit{permissible trajectory}, if for $\epsilon > 0$,
		\begin{align}
			v(t) = \frac{\left\Vert\tau_{N,d}(t)-\tau_{N,d}(t-1) \right\Vert_2}{\Ts}  & \leq \frac{\vmax}{2d\tan(\beta/2)} \quad \forall t > 0,\\
			|\theta_t - \thetaMU[t]|^2 + |\phi_t &- \phiMU[t]|^2 > \epsilon^2. \label{eq:angleBoundDef}
		\end{align}
		We use $\mathcal{T}_{N,d, \epsilon}$ to denote the set of all permissible trajectories.
	\end{definition}
	The parameter $\epsilon$ in \eqref{eq:angleBoundDef} characterizes the minimum permissible angular distance between the RX and the UAV, with respect to the TX. The constraint in \eqref{eq:angleBoundDef} prevents the UAV from blocking the LoS path between the TX and the RX.
	
	We now formulate the discrete trajectory optimization problem. The eavesdropper first computes the $q$-bit quantized beamformer $\tilde{\mathbf{F}}_t$ corresponding to the RX for all $t$. Then, the function $\mathcal{C}_t(\tilde{\mathbf{F}}_t, \tau(t))$ is evaluated over a discrete time grid. Finally, the optimal trajectory $\tau_{{N},d,\epsilon}^*$ can be defined as 
	\begin{align}
		\tau_{{N},d,\epsilon}^* := \argmin_{\tau \in \mathcal{T}_{{N},d,\epsilon}} &\sum\limits_{t = 0}^{(N-1)\Ts} \mathcal{C}_t(\tilde{\mathbf{F}}_t, \tau(t)). \label{eq:trajectoryOptimizationDiscrete}
	\end{align}
	The problem in \eqref{eq:trajectoryOptimizationDiscrete} finds an optimal trajectory from a set of permissible trajectories that maximizes the total secrecy rate over time $T$.
	
	\begin{algorithm}[tb]
		\begin{algorithmic}[1]
			\STATE{Initialize array $H$ as $H(\mathbf{s}) = 0$ for all $\mathbf{s}$}
			\FOR{$n = N-1, N-2\ldots,1$}
			\STATE{$H(\mathbf{s}) \leftarrow \max_{\mathbf{s}': (\mathbf{s},\mathbf{s}')\in\mathcal{A}} R(\mathbf{s}') + H^*(\mathbf{s'})\; \forall \mathbf{s}=(u,v,n)$}
			\ENDFOR
			\STATE{\textbf{Output:} A trajectory $\tau$ of length $N$, such that $\tau(0) = \argmax_{\mathbf{s}=(u,v,0)} H^*(\mathbf{s})$ and \\$\tau(t+1) = \argmax_{(\mathbf{s},\mathbf{s}')\in\mathcal{A}, \mathbf{s}_t = \tau(t)} \; R(\mathbf{s}') + H^*(\mathbf{s'})$.}
		\end{algorithmic}
		\caption[caption]{Value function estimation and optimal trajectory}\label{alg:Algo1}
	\end{algorithm}
	
	We solve the optimization problem in \eqref{eq:trajectoryOptimizationDiscrete} using a dynamic programming-based trajectory search. For that, we first define the state space, actions and reward as follows:
	\begin{enumerate}
		\item \textit{State:} The state of the UAV at time index $t$ is given by $\mathbf{s} = (u,v,t)$ where $(u,v) \in [-1,1]^2$ and $t\in \set{N}$. We also define the state at time $t$ as $\mathbf{s}_t = (u,v)$. 
		We use a discrete $G\times G$ spatial grid to represent the coordinates $(u,v) \in \{-1+2i/G: i \in \set{G}\}^2$.
		\item \textit{Action:} An action $a_t = (\mathbf{s}, \mathbf{s}')$ at time $t$ is defined as the transition from state $\mathbf{s} = (u,v,t)$ to $\mathbf{s}' = (u',v',t+1)$. 
		An action $a_t = (\mathbf{s}, \mathbf{s}')$ is a valid action if there exists a permissible trajectory $\tau \in \mathcal{T}_{N,d, \epsilon}$ that makes a transition from state $\mathbf{s}$ to $\mathbf{s}'$. 
		We denote the set of all valid actions by $\mathcal{A}$.
		\item \textit{Reward:} As the goal of the eavesdropper is to minimize \eqref{eq:secrecyRate}, we define the reward $R$ associated with an action $a_t = (\mathbf{s},\mathbf{s}')$ as 
		\begin{equation}R(a_t) = \log\left(1+\frac{P_{r_{t+1}}}{\sigma^2} \left\vert\frob{\mathbf{V}(\theta_{t+1},\phi_{t+1}) }{ \tilde{\mathbf{F}}_{t+1} }\right\vert^2\right),\end{equation}
		where $(r_{t+1},\theta_{t+1},\phi_{t+1}) = S_1(S_2(\mathbf{s}'))$.
		Since the definition of the reward solely depends on the next state, we denote $R(a_t) = R(\mathbf{s}')$ where $a_t = (\mathbf{s},\mathbf{s}')$.
	\end{enumerate}
	
	We now describe an adaption of dynamic programming called value iteration to solve \eqref{eq:trajectoryOptimizationDiscrete} \cite{SutBar:Reinforcement-learning:-An-introduction:18}. 
	The value function is defined as 
	\begin{align}
		H^*(\mathbf{s}) = \max_{\mathbf{s}':(\mathbf{s},\mathbf{s}') \in \mathcal{A}} \left[R(\mathbf{s}')+H^*(\mathbf{s}')\right].
	\end{align}
	An algorithm to estimate the value function is given in \algref{alg:Algo1}. Then, the optimal sequence of states that maximizes the reward, equivalently the optimal trajectory, is found using \algref{alg:Algo1}. We discuss the performance of the proposed trajectory search algorithm in \secref{sec:numerical}.
	
	We would like to highlight that our trajectory optimization algorithm requires the knowledge of the sequence of standard beamformers, i.e., $\{\tilde{\mathbf{F}}_t\}^T_{t=0}$, which can be computed from the trajectory of the RX. Furthermore, in a V2I system, the trajectory of the RX can be estimated based on the traffic geometry and vehicle dynamics. 
	Although the design of sophisticated real-time attacks that incorporate additional mechanical constraints such as the acceleration and power of the UAV is an interesting research direction, it is not within the scope of this work. 
	
	
	\section{Numerical results}\label{sec:numerical}
	In this section, we show the severity of the proposed attack and the benefit of the proposed CSB defense. Specifically, we first discuss the \smi achieved by CSB defense compared to the benchmark \dm-based technique, ASM \cite{ValLozHea:Antenna-Subset-Modulation:13}. 
	We then show the severity of the \textit{AirSpy} attack on a V2I TX, and explain the benefits of using CSB in terms of \ser against such an attack. 
	
	\subsection{Performance of the defense technique}
	In this part, we compare the CSB technique with ASM in terms of the \smi. 
	To this end, we consider a $16 \times 1$ linear phased antenna array at the TX and the use of the QPSK modulation. We consider an RX located at 25$^\circ$ with respect to the broadside angle of TX array. We plot the SMI for different angular positions of the eavesdropper located at the same radial distance from TX as the RX. 
	We denote the ASM technique by ASM-$c$ where $c$ denotes the fraction of active antennas at the TX.
	
	
	\begin{figure}[tb]
		\centering        
		\subfigure[1-bit phased array \label{fig:DefenseSER1D1bit}]{
			\includegraphics[width=0.45\columnwidth]{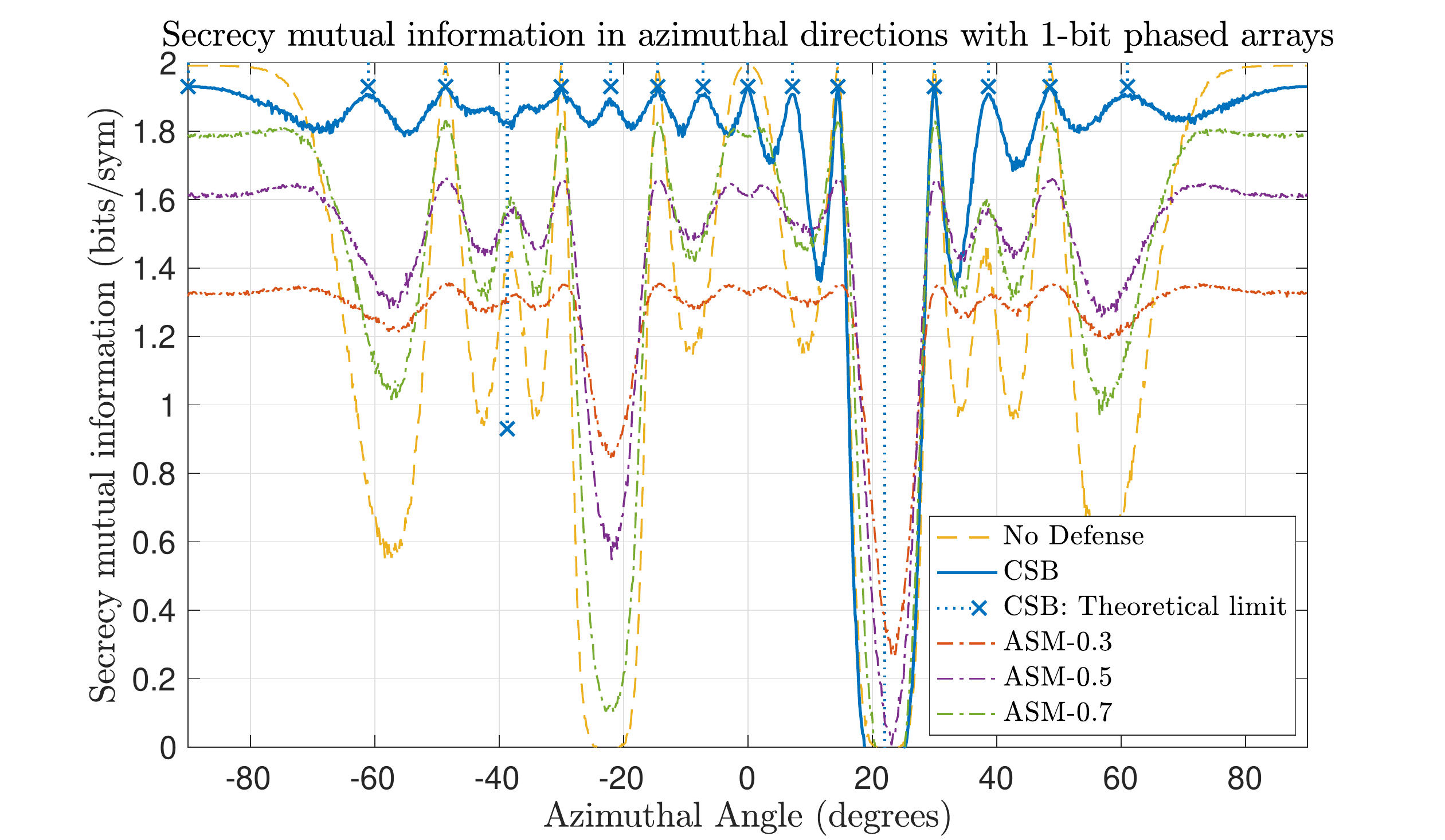}
		}
		\subfigure[2-bit phased array \label{fig:DefenseSER1D2bit}]{
			\includegraphics[width=0.45\columnwidth]{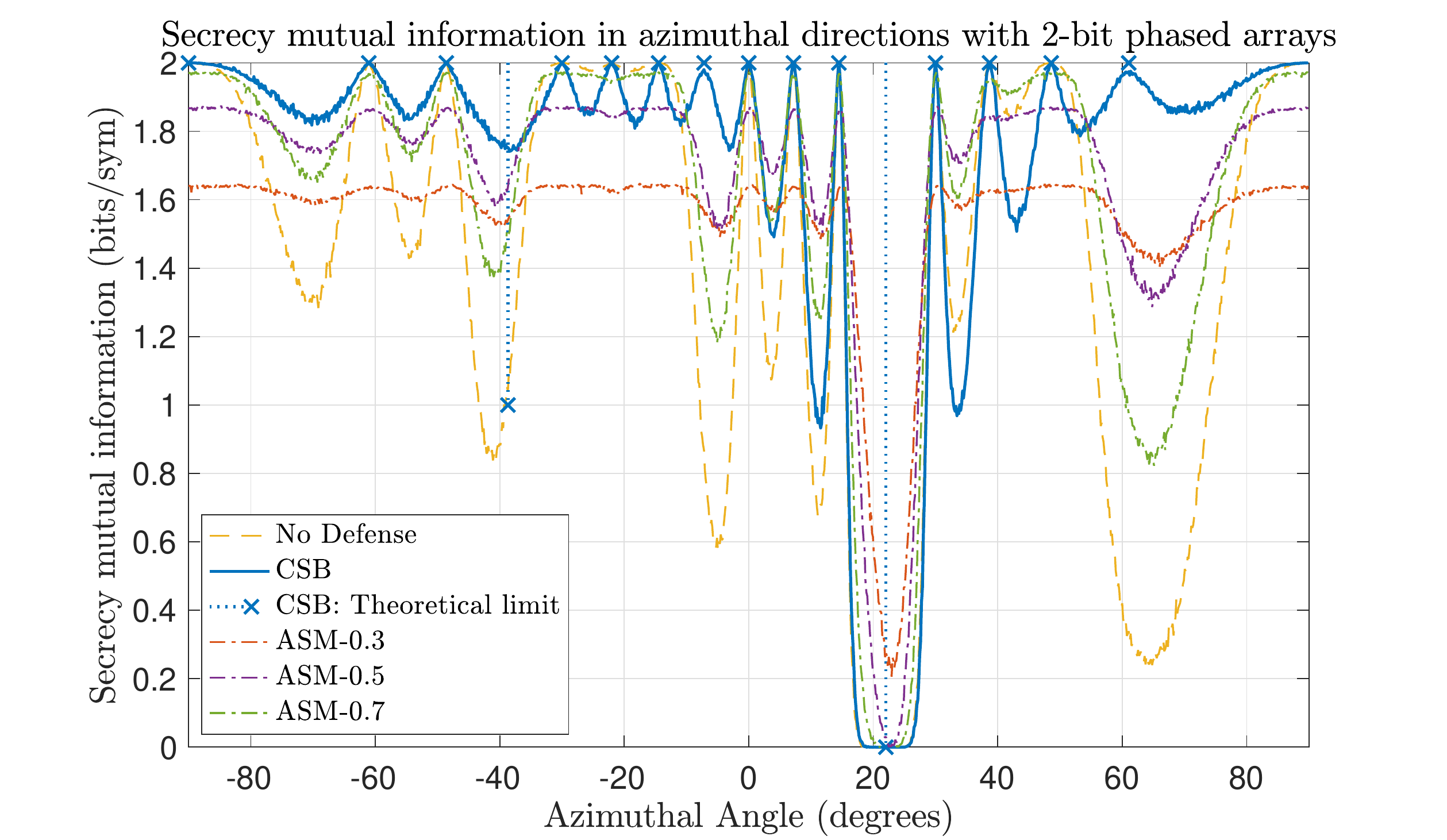}
		}
		\caption{\smi along different directions when the RX is at $25^\circ$ with respect to the boresight of the TX array.
			CSB defense achieves a large SMI as it preserves the \snr at the RX and induces \apn along the other directions. The theoretical SMI shown for on-grid positions is derived from \lemmaref{lemma:lemma3}. The theoretical SMI of 1 bit/sym at -38$^\circ$ corresponds to $g_t = 8$ as discussed in \exref{ex:ex1}.}
		\label{fig:DefenseSER1D}
	\end{figure}
	
	In \figref{fig:DefenseSER1D}, we show the numerically estimated SMI of CSB defense, and ASM defense with 0.3, 0.5 and 0.7 fraction of active antennas. 
	We notice that ASM performs poor along the directions of the energy leakage. This is due to the fact that the \AN induced by ASM is small when compared to the RF signal leakage with low-resolution phased arrays.  Furthermore, ASM defense also suffers from lower received power at the RX under the common per-antenna power constraint. 
	In contrast, CSB defense achieves better SMI as compared to ASM. We also plot the theoretical mutual information transfer at high SNR for on-grid positions of the eavesdropper as characterized in \lemmaref{lemma:lemma3}. Note that the theoretical SMI for CSB along $-38.6^\circ$ corresponds to the position of the eavesdropper such that $g_t = 8$ as discussed in \exref{ex:ex1}. Due to lower energy leakage, however, the SMI along that direction is still higher than ASM. 
	
	\subsection{Severity of AirSpy attack}
	In this part, we numerically show the severity of the proposed attack. We first provide the trajectory of the UAV calculated with our trajectory design algorithm. Then, we study the secrecy rate of the system corresponding to the designed trajectory.
	
	We consider a downlink V2I scenario, shown in \figref{fig:UAVPlane}, where the TX is equipped with a planar mmWave phased array with $16\times16$ elements. The TX array is located at $h=8\, \mathrm{m}$ above the ground and is tilted downward by $15^\circ$. A vehicular RX travels on a straight lane at a distance of $\ell=3$ m from the TX at a speed of $20$ m/s. We assume that the RX is in a connected mode with this TX when the transceiver distance along the $y-$dimension is within $10$ m, i.e., $ y_t \in [-10, 10]$. As the vehicle moves at $20$ m/s, the RX is connected to the TX for $1$ second. We call this $1$ second duration as \textit{an episode}.
	
\begin{figure}[tb]
		\centering        
		\subfigure[Trajectory of the eavesdropper\label{fig:AttackTrajectory}]{
			\includegraphics[width=0.4\columnwidth,height=2in,keepaspectratio]{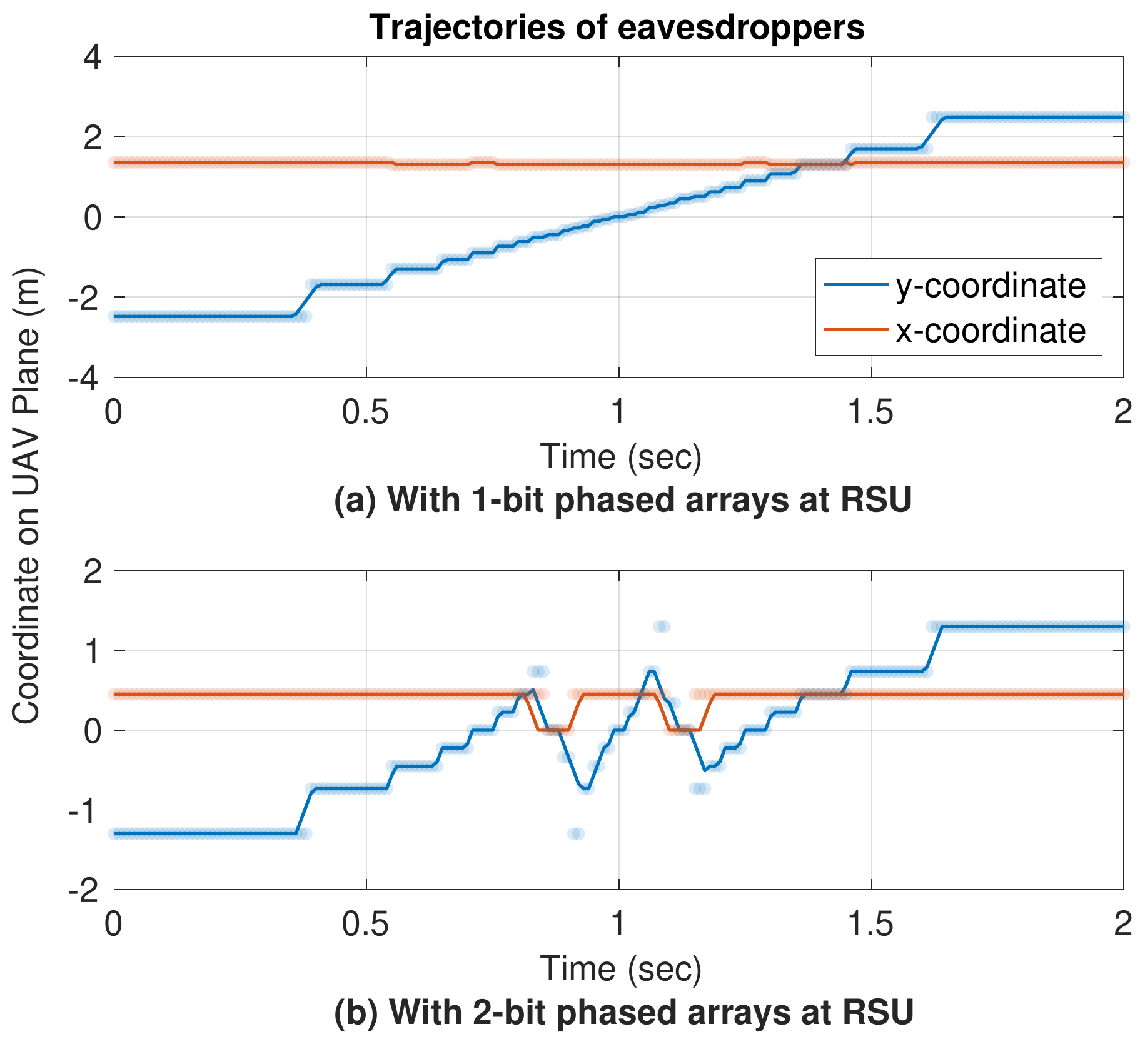}
		}\quad\quad\quad
		\subfigure[Secrecy rate over an episode\label{fig:AttackSecrecyRate}]{
			\includegraphics[width=0.4\columnwidth,height=2in,keepaspectratio]{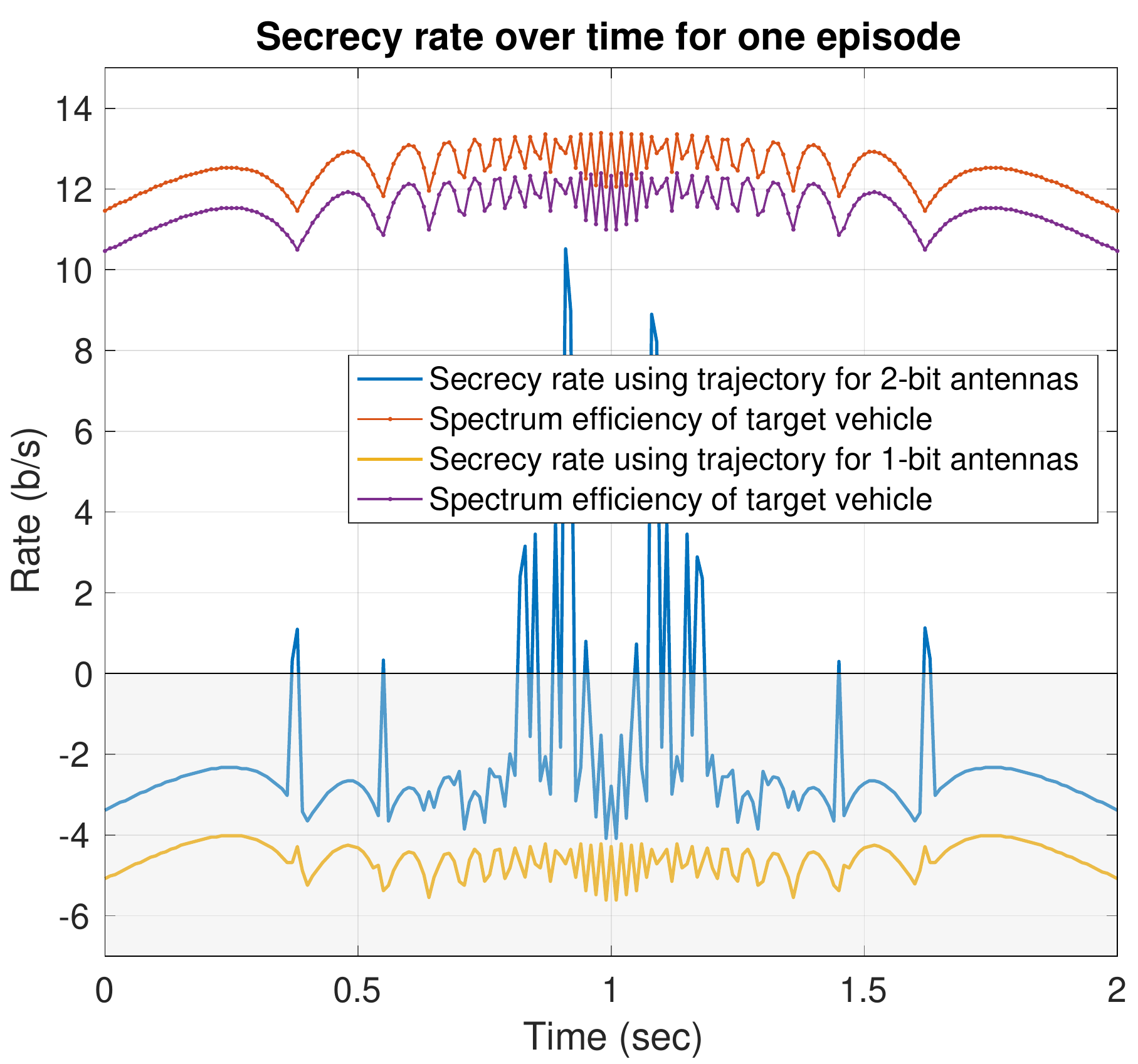}
		}
		\caption{The figure depicts attacks using \textit{AirSpy}. In (a), we show the optimal trajectory of the eavesdropper on the UAV plane, and the strongest sidelobe with dots. For one-bit phased arrays, the eavesdropper just tracks the strongest sidelobe. For 2-bit phased arrays, however, the eavesdropper follows a different path to avoid sudden transitions that arise when tracing the strongest side-lobe. This is because such sudden transitions are mechanically infeasible. The evolution of the secrecy rate over an episode is illustrated in (b). AirSpy is a good attack that substantially reduces the secrecy rate in low resolution phased array systems.}
		\label{fig:Attack}
	\end{figure}

	We assume that the UAV eavesdropper traverses on a plane at a distance $d = 1$ m from the TX array. For the simulation, we consider a bounded region of the plane such that the angle subtended by the region at the center of TX antenna array is $\beta = 160^\circ$. We limit the speed of the UAV to $17$ m/s \cite{Gre:Announcing-the-Dreamer-Drone:}. In this setting, we first plot the eavesdropping trajectory designed using our dynamic programming-based algorithm when the RX moves from point $(3,-10,8)$ to $(3,10,8)$ in an episode. The trajectories derived for attacks on $1$-bit and $2$-bit phased arrays are shown in \figref{fig:AttackTrajectory}.
	
	We notice that the optimal trajectory for eavesdropping on a one-bit phased array TX is consistent with the analytical solution derived in Appendix \ref{appendix:onebitreflection}. The solution can be explained from the observation that the beams generated with a one-bit phased array are mirror symmetric about the boresight direction. In case of 2-bit phased arrays, however, the optimal eavesdropping trajectory derived with our method exhibits an interesting phenomena. The UAV diverges from the direction of the strongest side-lobe at about $0.8$ seconds and $1.2$ seconds. This divergence is important to minimize the sum secrecy rate over an episode. Such a change results in better eavesdropping than a feasible greedy trajectory that simply follows the strongest sidelobe. We illustrate this observation using a video that is available on our website \cite{kkpatel1-CSBAirSpy}.
	
	In \figref{fig:AttackSecrecyRate}, we show the evolution of the secrecy rate as the eavesdropper follows the trajectory shown in \figref{fig:AttackTrajectory} during one episode. 
	The secrecy rate when using one-bit phased arrays at TX is consistently $0$ because the energy received at the UAV eavesdropper is higher than the energy received at the RX. This is because the UAV eavesdropper is closer to the TX than the RX. The secrecy rate using the trajectory designed for 2-bit phased arrays at the TX is also below $0$ for the same reason, except during the time when the eavesdropper deviates from the path traced out by the strongest side-lobe.
	
	In both the one-bit and the two-bit scenarios, the rate at the eavesdropper is significantly higher than the rate at the RX. 
	In such a case, any defense strategy that slightly reduces the leaked RF signals does not help in minimizing the secrecy rate. Furthermore, strategies that null the leaked RF signal in a particular direction are also not useful. This is because a mobile eavesdropper can optimize its trajectory in the new setup to track the other side-lobes. Therefore, any defense technique that reduces the energy leakage cannot tackle the issue of eavesdropping with a mobile eavesdropper. Our CSB defense corrupts the phase of the symbols along the directions other than the direction of the RX, instead of reducing the energy leakage.
	
	\noindent\textbf{Remark: }Although the secrecy rate is a non-negative quantity, we plot negative values in \figref{fig:Attack} to show the large difference between the rates at the RX and the eavesdropper over an episode.
	
	\subsection{Defense against AirSpy}
	We describe the benefits of using CSB defense over ASM in a low-resolution phased array under the AirSpy attack. We use a system setup similar to the one used to analyze the attack. 
	For the simulation of CSB and ASM defense, we consider both the RX and the eavesdropper perform perfect synchronization and we only focus on the performance during the data transmission. 
	Additionally, we consider that the TX corrects the phase change as characterized in \lemmaref{lemma:lemma1} when the RX is along an on-grid direction or an off-grid direction. Since the nearest on-grid direction associated with the RX is known to the TX in the form of the beam selected from the DFT codebook, our defense method does not require additional information to maintain the communication performance at the RX. Note that the phase change due to circulant shifts characterized in \lemmaref{lemma:lemma1} is only valid along the on-grid directions. We will show using simulations that the phase correction based on nearest on-grid direction still maintains the performance at the RX along the off-grid directions. Furthermore, we assume a standard receiver to calculate the \ser. 
	
	\begin{figure}[tb]
		\centering
		\subfigure[\centering \ser at the eavesdropper with \snr at the RX \label{fig:SERVsSNR}]{
			\includegraphics[width=0.3\columnwidth]{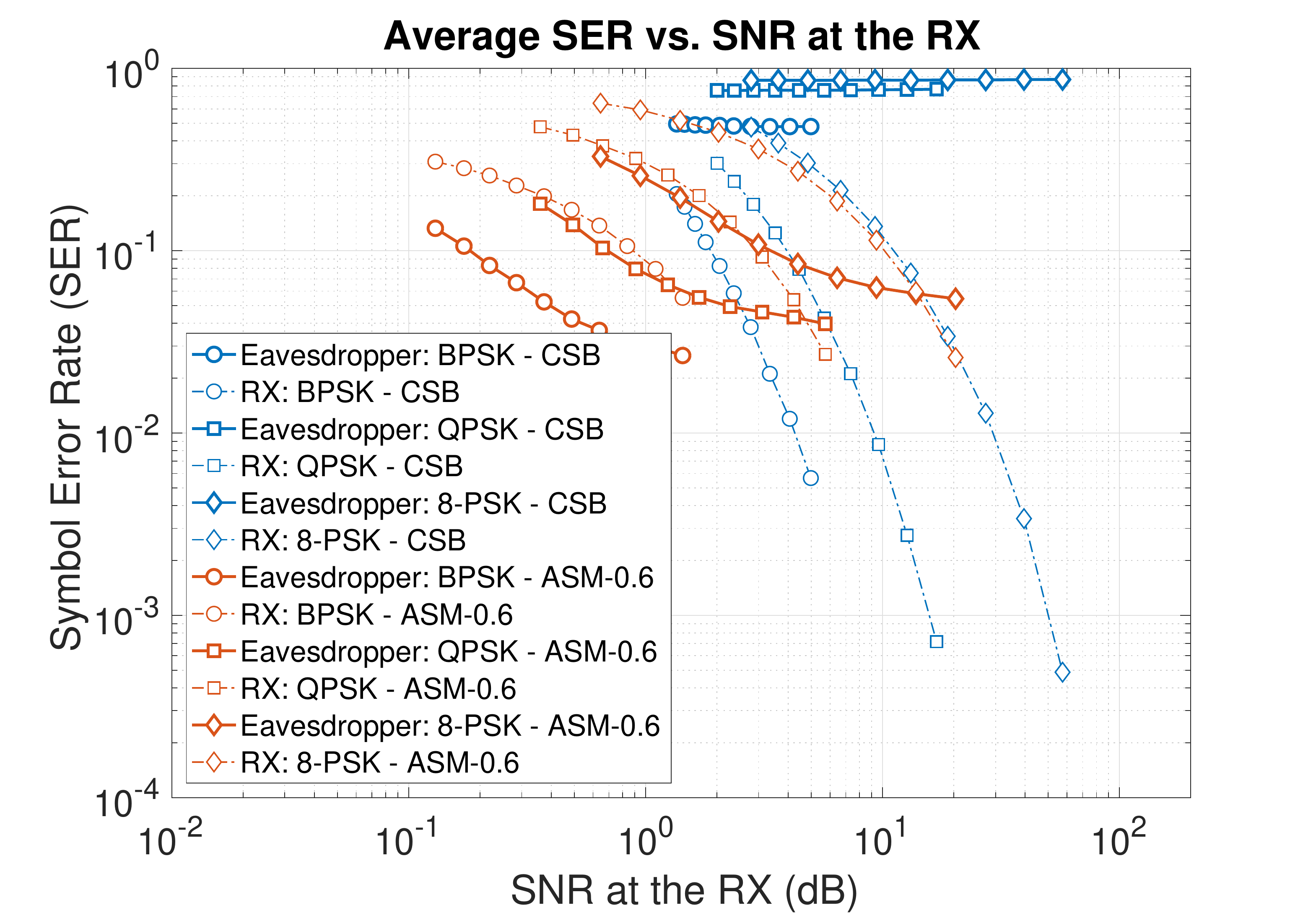}
		}
		\subfigure[\centering \ser at the eavesdropper and the RX for different ASM parameters \label{fig:SER_vs_ASMFrac}]{
			\includegraphics[width=0.3\columnwidth]{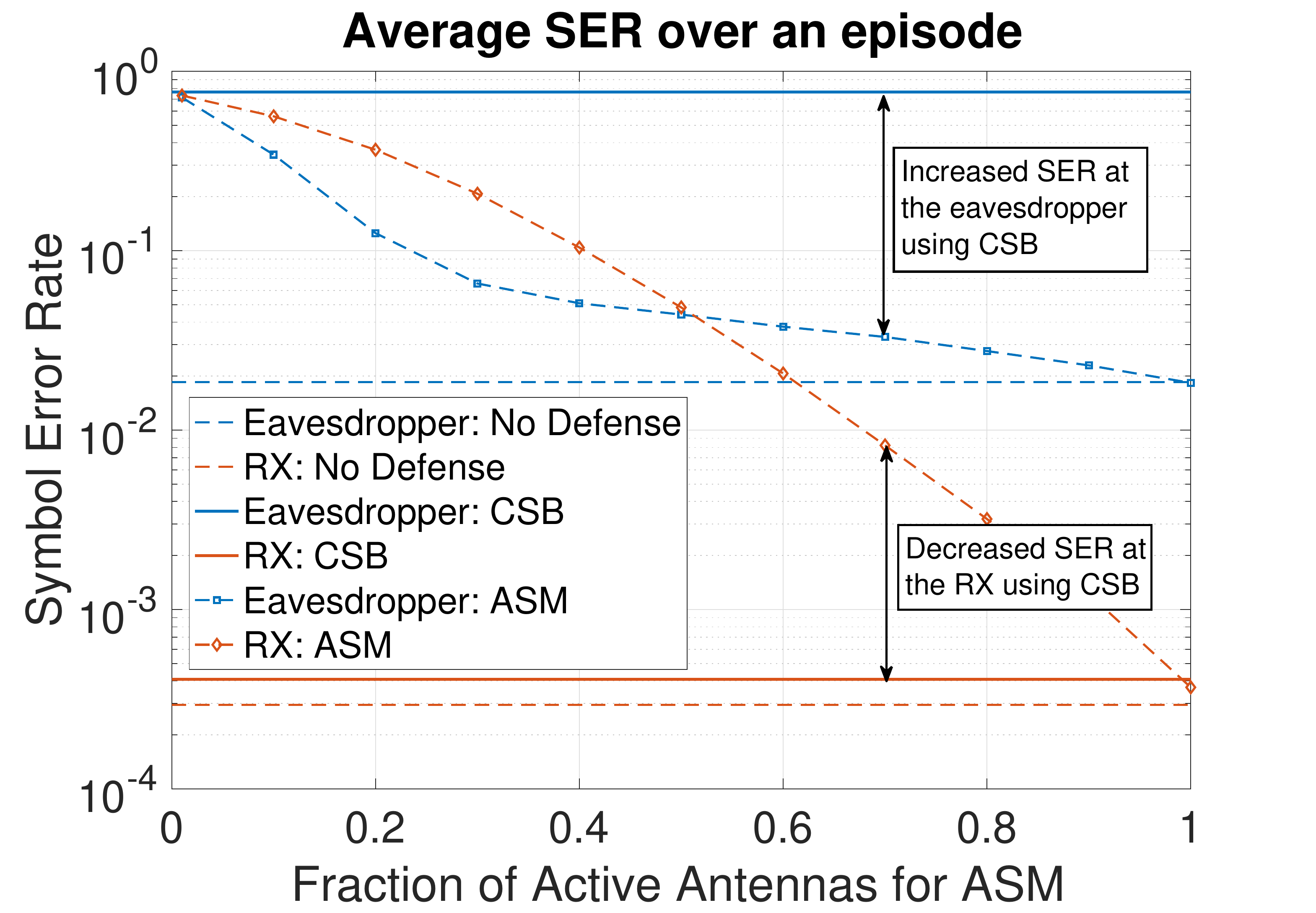}
		}
		\subfigure[\centering \snr at the RX for different ASM parameters \label{fig:SNR_vs_ASMFrac}]{
			\includegraphics[width=0.3\columnwidth]{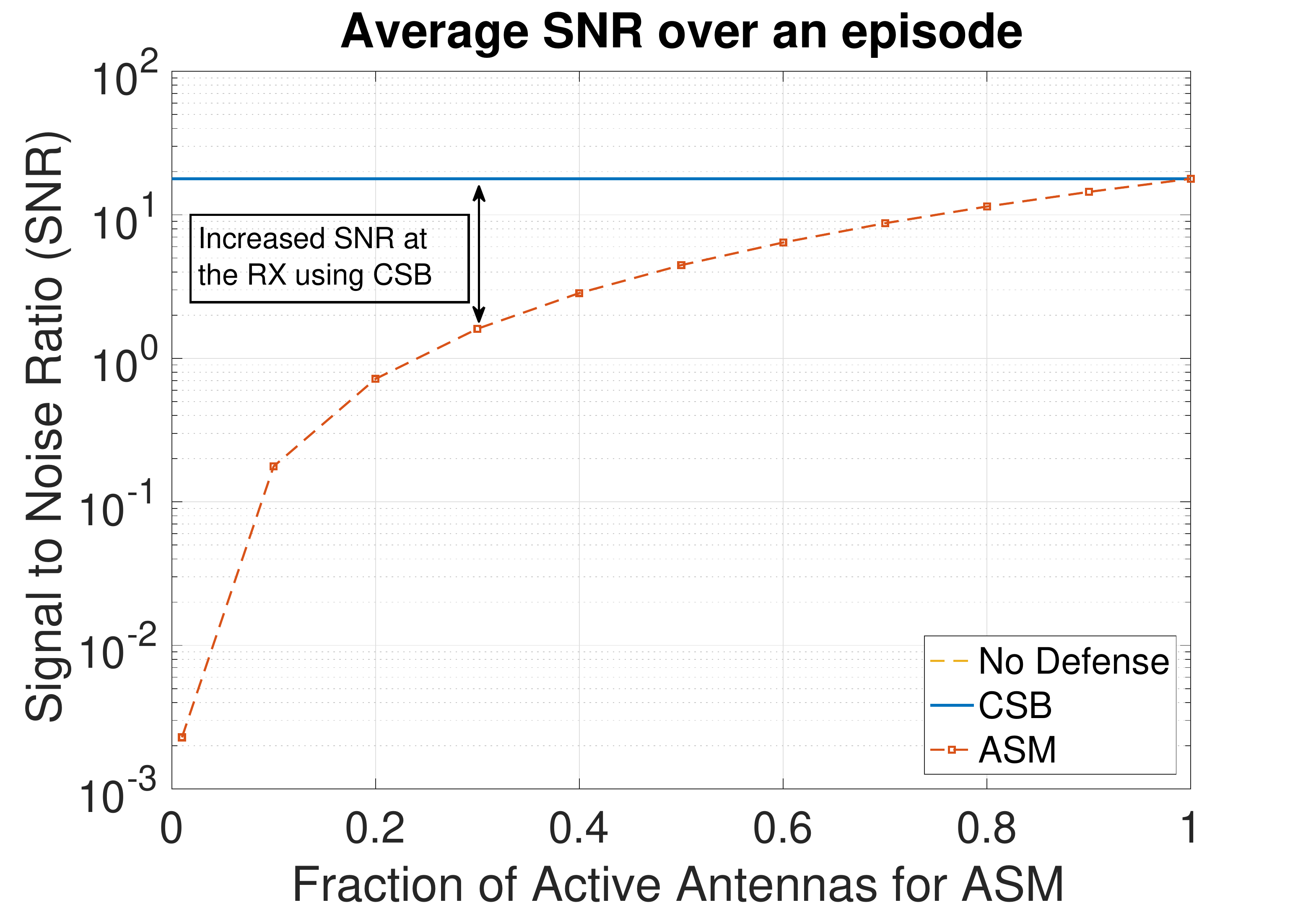}
		}
		\caption{The plots show the \ser and \snr performance of CSB defense as compared to ASM when the TX with 2-bit phased array is under the AirSpy attack. ASM provides lower \ser to the eavesdropper as compared to CSB. CSB also provides higher \snr at the RX as compared to ASM.}
		\label{fig:DefenseMobile}
	\end{figure}
	
	In \figref{fig:SERVsSNR}, we show the average \ser at the RX and the eavesdropper as the function of the \snr received at the RX. Note that the \ser at the RX is higher than the \ser at the eavesdropper when using ASM-0.6 for the defense. This is due to two reasons. First, the received power at the eavesdropper is higher than the received \snr at the RX as the TX-eavesdropper distance is much smaller than the TX-RX distance. Second, the \AN induced by ASM which adds to the noise at the eavesdropper is not sufficient enough to perturb the constellation at the eavesdropper. Thus, the effective signal power received at the eavesdropper due to the signal leakage from the low-resolution phased arrays is higher than the \AN induced by ASM. In contrast, CSB defense scrambles the phase of the signal along the directions other than that of the RX, thus, corrupting the signal irrespective of the signal power.
	
	In \figref{fig:SER_vs_ASMFrac}, we show the average \ser at the eavesdropper and the RX for different ASM parameter $c$. The \ser at the eavesdropper when using CSB defense is higher than ASM defense for any parameter $c$. Additionally, the \ser at the RX is also consistently lower when using CSB as compared to using ASM. It can also be observed from \figref{fig:SNR_vs_ASMFrac} that the use of CSB defense also provides an increased \snr at the RX when compared to ASM. From \figref{fig:SER_vs_ASMFrac} and \figref{fig:SNR_vs_ASMFrac}, we can conclude that CSB achieves a large \ser at the eavesdropper, while the \ser and the \snr at the RX is maintained without any significant degradation from the standard case.

	\section{Conclusion}
	In this paper, we developed a directional modulation-based beamformer design technique called \textit{CSB}, to defend against an eavesdropping attack on low-resolution phased arrays. The proposed CSB defense applies random circulant shifts of the low resolution beamformer to scramble the phase of the received symbol in the unintended directions. As a result, CSB blinds an eavesdropper that taps the leaked RF signals. We characterized the phase ambiguity introduced at the eavesdropper and derived the secrecy mutual information. 
	We also designed an experiment on an mmWave testbed using 60 GHz phased arrays and showed that circulantly shifting a beamformer induces different but predictable phase shifts along different directions. The predictability of the phase shifts allows the TX to adjust the phase of the transmitted symbol to maintain the communication between the TX and the RX. 
	Finally, we developed an eavesdropping attack for low-resolution phased arrays in a V2I network and evaluated the performance of CSB under such an attack. Our results indicate that CSB achieves a better defense than similar state-of-the-art benchmark techniques.
	
	\bibliographystyle{IEEEtran}
	\bibliography{main.bib}
	
	\appendix
	\subsection{Proof that the beams with one-bit phased arrays are symmetric about the boresight}\label{appendix:onebitreflection}
	We show that a one-bit rectangular phased antenna array generates an unintended beam along a direction that is mirror symmetric to the desired direction about the boresight. We use $\tilde{\mathbf{F}}_t$ to denote a one-bit beamformer which maximizes $|\frob{\mathbf{V}(\thetaMU[t],\phiMU[t])}{\tilde{\mathbf{F}}_{t}}|^2$, i.e., the energy of the beam in the direction of the RX. We observe that the entries of the one-bit beamformer are $\pm1/ \Ntx$.  The energy leakage in the mirror symmetric direction to the RX, i.e., $(-\thetaMU[t],-\phiMU[t])$, is determined by 
	$\left\vert\frob{\mathbf{V}(-\thetaMU[t],-\phiMU[t])}{\tilde{\mathbf{F}}_{t}}\right\vert^2$. 
	This is the same as $ \left\vert\frob{\bar{\mathbf{V}}(\thetaMU[t],\phiMU[t])}{\tilde{\mathbf{F}}_{t}}\right\vert^2
	$, by the property that $\bar{\mathbf{V}}(-\theta, -\phi)=\mathbf{V}(\theta, \phi)$. Now, we observe that $(\bar{\tilde{\mathbf{F}}}_{t})=\tilde{\mathbf{F}}_{t}$ as the one-bit beamformer has real entries. As a result,
	\begin{align}
		\left\vert\frob{\mathbf{V}(-\thetaMU[t],-\phiMU[t])}{\tilde{\mathbf{F}}_{t}}\right\vert^2&=\left\vert\frob{\bar{\mathbf{V}}(\thetaMU[t],\phiMU[t])}{\bar{\tilde{\mathbf{F}}}_{t}}\right\vert^2=\left\vert\frob{\mathbf{V}(\thetaMU[t],\phiMU[t])}{\tilde{\mathbf{F}}_{t}}\right\vert^2.
	\end{align}
	Therefore, the beam pattern with a one-bit phased array has an equal amount of energy along the directions $(\thetaMU[t],\phiMU[t])$ and $(-\thetaMU[t],-\phiMU[t])$. 
	Due to this property, we observe that a reasonable eavesdropping strategy is one that traces the mirror-symmetric path corresponding to the RX.
	\label{LastPage}
\end{document}

